\documentclass[aps,pre,amsmath,amssymb,reprint,onecolumn,floatfix,longbibliography,notitlepage]{revtex4-1}

\usepackage{graphicx}
\usepackage{dcolumn}
\usepackage{bm}
\usepackage[section]{placeins}

\begin{document}

\preprint{AIP/123-QED}

\title{Assessing turbulence sensitivity using stochastic Monte Carlo analysis}

\author{K. Duraisamy, Anand A.}
\affiliation{Aerospace Engineering Department, University of Michigan, Ann Arbor 48109, USA}
\author{G. Iaccarino}%
\affiliation{Center for Turbulence Research, Stanford University, Stanford, California 94305, USA}


\begin{abstract}
Reynolds Averaged Navier Stokes (RANS) models represent the workhorse for studying turbulent
flows in industrial applications. Such single-point turbulence models have limitations in accounting for the influence of the non-local physics and flow history on the evolution of the turbulent flow. In this context, we investigate the sensitivity inherent to single-point models due to their characterization of the internal structure of homogeneous turbulent flows solely by the means of the Reynolds stresses. For a variety of mean flows, we study the prediction intervals engendered due to this coarse-grained description. The nature of this variability and its dependence on parameters such as the mean flow topology, the initial Reynolds stress tensor and the normalized strain rate is identified, analyzed and explicated. The ramifications of this variability on the formulation of classical RANS closure models are highlighted.
\end{abstract}

\maketitle

\section{Introduction \& Overview}
In spite of over a century of research, no analytical solutions for the equations governing turbulent flows are available. Furthermore, with the state of computational resources, an exact numerical resolution of the scales of motion encountered in turbulent flows in engineering applications is neither feasible, nor foreseen. Consequently, most industrial investigations have to resort to some degree of modeling while studying turbulent flows numerically. Reynolds Averaged Navier-Stokes models represent the pragmatic recourse for complex engineering flows, with a majority of simulations resorting to this avenue.  Notwithstanding their wide spread use, RANS models suffer from an inherent structural inability to replicate fundamental turbulence processes and specific flow phenomena. Over the past few decades, there has been a large increment in the utilization of computational fluid dynamics (CFD) based simulations in the engineering design process. However, for novel designs, critical decisions or certification purposes, such numerical results are always supplemented with experimental data. This is due to the lack of universal reliability of RANS model predictions for different flows and disparate quantities of interest, leading to limited trust in RANS models. In this regard, the quantification of such uncertainties is of fundamental importance to enhance the predictive capabilities of RANS simulations as an engineering tool. This exercise establishes the degree of confidence in the model's predictions of different quantities of interest, for different flows under diverse conditions. 

The sources of the epistemic uncertainty inherent to RANS models can be broadly divided into two categories: \textit{structural} and \textit{parameter} uncertainty. Structural uncertainty arises due to the inadequacy of the closure expression to represent the underlying physics in turbulent flows. Turbulence models are constitutive relations attempting to relate quantities of interest to flow parameters, using a variety of assumptions and simplifications based on physical intuition and observations. Often, the limited applicability and accuracy of these assumptions may introduce an appreciable degree of uncertainty in the predictions. For instance, numerous RANS closures rely on the Boussinesq hypothesis to relate the instantaneous Reynolds stress tensor to the local mean rate of strain. While this assumption is relatively useful for simple shear flows, such an approximation is unsuitable in more complex flows such as those involving streamline curvature, flow separation or three-dimensional effects. This assumption ignores the non-local nature and the memory of the turbulence phenomenon and leads to significant limitations in the modeling of flows of engineering relevance. In this context, an excellent investigation into the limitations of the Boussinesq hypothesis has been carried out in \citet{ling1}. Similarly, turbulence model expressions use a number of closure coefficients which are usually assumed to be constant and are determined by calibration against a database of benchmark flows. The appropriation of the best-possible values of these coefficients, and, their functional form introduces additional uncertainty in the model framework. This is termed as parameter uncertainty. In this regard, \citet{edelinga} have investigated the parameter variability in the $k-\epsilon$ turbulence model. Similarly, \citet{margheri} and \citet{edelingb} have focused on investigating the sensitivity of the model predictions on the values of the model coefficients.

A key source of model inadequacy is the coarse-grained description adopted in RANS models, wherein the state of the turbulent flow field is described solely by means of the Reynolds stresses. All classical turbulence models assume that the state of turbulence, and, its evolution can be completely and uniquely described by a finite set of local tensors. For instance, popular two-equation RANS closures assume that turbulence is completely described by the turbulent kinetic energy and the rate of dissipation \citep{joneslaunder}. Similarly, RANS models based on the second moment closure methodology assume this to be the rate of dissipation and the Reynolds stress tensor components \citep{mishra1}. While of engineering utility to obtain closure, this assumption severely limits the modeling basis and therefore, the flow physics that such models may replicate. In this regard, \citet{tf61} have exhibited that even augmenting the classical modeling basis with additional tensors, such as Dimensionality, Stropholysis, Circulicity, Inhomogeneity, etc, is unable to capture the flow dynamics arising due to mean rotation. Similarly, \citet{cambon2006} have proved that a satisfactory description of just the anisotropy in the componentiality of the turbulent flow field requires a minimal set of three tensors. While the evolution of the stochastic velocity field governed by the Navier Stokes equations is Markovian, at the single-point description the evolution of the Reynolds stresses is unclosed and non-Markovian \citep{kraichnan}. However, this restriction to a limited, finite basis obligates turbulence models to be Markovian. This attempt to model a non-Markovian process via a Markovian closure has its drawbacks. For instance, as a ramification of this coarse-graining, such a framework implies that initially two-component turbulence will remain at the 2C limit, irrespective of the applied mean gradient \citep{rng}.

This coarse-grained description leads to problems in predictive fidelity and realizability, where the model is unable to replicate the true dynamics of turbulence. A more subtle
ramification is that of variability. Ignoring the history of the flow implies that specifying
the macro-state leads to a unique specification of the micro-states: specifically, that all
turbulent flows with the same Reynolds stresses will behave identically under similar
conditions. However, at this level of description, the turbulence modeling problem is ill-posed in the sense of \citet{Hadamard} and evolution of the real flow is non-unique. In this investigation, we exhibit that there is a broad range of possible flow evolution, often with diametric behavior. RANS models, being deterministic, may (at best) replicate a single evolution from this range. Consequently, at this level of description, the disparity between the broad range of possible evolutions for the real flow and the unique evolution predicted by the RANS model introduces epistemic uncertainty in the model’s predictions.

The variability arising due to this coarse grained description of the flow is common to all single-point closures and is independent of the modeling paradigm, the closure expression or the coefficients. Thus, the best possible single-point closure, with an ideal closure expression and an optimal set of coefficients, would still have this degree of uncertainty in its predictions. Its universal nature adds to the importance of this source of uncertainty.

Furthermore, this variability can add to the parameter uncertainty in the classical turbulence modeling framework. At present, most established turbulence models subscribe to a completely deterministic approach towards forecasting turbulence evolution. The parameters in such models are chosen to optimize agreement with a small set of canonical flows wherein high fidelity numerical or experimental data may be available. Such tuning of the parameters to \textit{single realizations} of the flow ignores the possibility of any variability in flow evolution at this level of description. Such a point estimate can still be optimal, in the root mean square error sense, if it agrees with the mean of the ensemble of possible realizations. However, it has been exhibited in \citet{epstein1969} that the evolution of this ensemble mean is different from any of the constituents of the ensemble. Thus, due to this coarse-grained description, the coefficient calibration against singleton realizations of benchmark flows may lead to greater parameter uncertainty and deficient estimates for these coefficients. 

The evolution of this uncertainty can be expressed in terms of a Liouville conservation equation for the growth of initial uncertainty, or a Fokker-Planck equation, if additional model inadequacies are also considered. Such uncertainty arising due to an inadequate description of the initial state arises in other fields such as meteorology and oceanography as well \citep{palmer2000, thompson1957}. For instance, in Numerical Weather Prediction (NWP), the initial uncertainty arises due to limited, sparse and low-resolution measurements describing the state of the atmosphere \citep{bauer2015}. In such practical applications, the distribution of the error in the description of the initial state cannot be estimated with sufficient accuracy to resort to analytical approaches such as the Liouville equation. In practice, these equations are solved using ensemble forecasting techniques. Herein, the deterministic governing equations are integrated for an ensemble of initial conditions, each consistent with uncertainties in the initial data. The perturbed initial conditions constitute a random sampling of the initial probability distribution of the uncertainty and thus, this methodology represents a stochastic Monte Carlo approach to solve the problem. In the recent past, ensemble forecasting has been utilized successfully for a host of commercial applications such as electricity generation, ship routeing, pollution modeling, disease prediction, crop yield modeling, besides others \citep{palmer2000,taylor2003}.

In this article, we seek to address the qualitative and quantitative aspects of the variability introduced into the RANS modeling problem due to the coarse grained description of the turbulent flow field. This involves determining the prediction intervals for the evolution of turbulence statistics for diverse flows, under varying conditions. Additionally, the nature of this uncertainty is evaluated with respect to its dependence on the state of the initial state of the Reynolds stress tensor, the applied mean gradient and the normalized strain rate parameter.

Further, this investigation provides physics based guidance for structural uncertainty quantification using eigenvalue perturbations. Recently, a physics-based, non-parametric approach to estimate the model-form uncertainties via injecting perturbations in the Reynolds stresses has been developed by \citet{emory1}. This has been applied to engineering problems with considerable success \citep{gorle1,gorle2,gorle3,xiao,mishra2016,mishra5}. We outline the manner in which the results of this study can be used to provide guidance in determining the magnitude of the perturbations to the turbulent kinetic energy.

\section{Mathematical Details}
In this section, we outline the mathematical rationale underlying this investigation. After an introduction to the nature of the uncertainty, three specific aspects are discussed: the truth model, the representation of the initial uncertainty and the set of applied mean flows considered herein.

The deterministic equations governing the time evolution of a system can be expressed schematically as
\begin{equation}\label{eq:toyequation}
\dot{x}=f(x),
\end{equation}
where $x$ is an n-dimensional vector, describing the instantaneous state of the flow. For homogeneous turbulent flows, $x$ would pertain to the components of the fluctuating velocity field and pressure, and, $f$ is a non-linear function. In the $N$-dimensional phase space for the components of $x$, $f$ defines a trajectory for the evolution of the initial state. If there exists an uncertainty in the specification of the initial state, it may be represented by the probability distribution function, $\rho (x, \tau=0)$. Consequently, over a volume in phase space, $V$, the probability that the true initial state lies in this volume is given by $P(x^{initial}_{truth} \in V)=\int_{V} \rho (x, \tau=0) dV$. The evolution of this distribution is given by the Liouville equation:
\begin{equation}\label{eq:liouville}
\frac{\partial \rho}{\partial t}=-\frac{\partial}{\partial x} (\rho\dot{x}),
\end{equation}
where $\dot{x}$ is defined by equation~\ref{eq:toyequation}. If we consider there to be additional uncertainties due to our limitations in the model used, then equation~\ref{eq:toyequation} can be re-expressed as
\begin{equation}
\dot{x}=f_1(x)+f_2(x,\alpha(x)),
\end{equation}
where $f_1$ represents the terms in the equations of motion that are completely resolved in the modeling procedure and $f_2$ is a parametric representation of the dynamics of the unresolved scales \citep{williams2005}. If the model uncertainties can be represented by a zero-mean, Gaussian white noise term, then equation~\ref{eq:liouville} corresponds to the form of a Fokker-Planck equation. 

Consequently, we observe that the evolution of this initial uncertainty can be expressed in terms of a Liouville conservation equation, or a Fokker-Planck equation, if model uncertainties are also considered. In NWP, these are solved via \textit{ensemble forecasting} \citep{zoltan}. Herein, to estimate the non-linear evolution of the probability density function of the uncertainty, Monte-Carlo techniques are utilized. The probability distribution is sampled form at an initial time and the deterministic equations are used to evolve these sampled initial states. Herein, the initial sample is referred to as the ensemble and individual elements of this collection as ensemble members.  

In this investigation, we utilize the Interacting Particle Representation Method (IPRM) \citep{iprm2}. The IPRM has a viscoelastic character for slow deformations and is exact for rapid deformations, wherein it has an elastic character. The turbulent field is represented as a collection of idealized building-blocks for its internal structure. These hypothetical particles are characterized by a velocity vector, $\mathbf{V}$; a vorticity vector, $\mathbf{W}$; a stream function vector, $\mathbf{S}$; pressure, $P$ and a gradient vector, $\mathbf{N}$. Quantities of interest can be expressed in terms of these properties of the ensemble. For instance, the one-point statistic of the Reynolds stress, $R_{ij}$, can be expressed as: $R_{ij}=\langle V_iV_j \rangle$, where $\langle \quad \rangle$, represents the average over the ensemble of particles. Similarly, the dimensionality of the turbulent flow can be expressed as $D_{ij}= \langle S_nS_nN_iN_j \rangle$ and the circulicity as $F_{ij}= \langle S_iS_jN_nN_n \rangle$. Additionally, for homogeneous turbulence, these tensors can be normalized using the turbulent kinetic energy, $k$, to give:
\begin{equation}
r_{ij}+d_{ij}+f_{ij}=\delta_{ij},
\end{equation}
where $r_{ij}=\frac{R_{ij}}{2k}$, $d_{ij}=\frac{D_{ij}}{2k}$ and $f_{ij}=\frac{F_{ij}}{2k}$.

For reasons of computational efficiency and stability, we utilize the cluster averaged formulation of the IPRM. Herein, the particles are grouped in clusters having a common gradient vector. The averaging over particles is carried out in two discrete steps. First, an averaging is done over particles in a cluster, followed by an averaging over all clusters. The statistics resulting from the first step are moments conditioned upon the gradient vector, given by $R^{|N}_{ij}=\langle V_iV_j|\mathbf{N} \rangle$, $D^{|N}_{ij}= \langle S_nS_nN_iN_j|\mathbf{N}  \rangle$ and $F^{|N}_{ij}= \langle S_iS_jN_nN_n|\mathbf{N}  \rangle$. 

For this IPRM formalism, the evolution equations for the unit gradient vector, $n_i=\frac{N_i}{N}$, and the conditional Reynolds stress tensor, $R^{|n}_{ki}$, are given by:
\begin{equation}
\begin{split}
\dot{n_i}=&-G^n_{ki}n_k+G^n_{kr}n_kn_rn_i,\\
\dot{R}^{|n}_{ij}=&-G^{v}_{ik}R^{|n}_{kj}-G^{v}_{jk}R^{|n}_{ki}+[G^{v}_{km}+G^n_{km}](R^{|n}_{im}n_kn_j+R^{|n}_{jm}n_kn_i)-C_r[2R^{|n}_{ij}-R^{|n}_{kk}(\delta_{ij}-n_in_j)]
\end{split}
\end{equation}
Herein, $G^n_{ij}$ and $G^v_{ij}$ denote the effective gradients for the evolution of the particle gradient vector and the particle velocity vector, given by
\begin{equation}
G^n_{ij}=G_{ij}+\frac{C^n}{\tau}r_{ik}d_{kj}, \quad G^v_{ij}=G_{ij}+\frac{C^v}{\tau}r_{ik}d_{kj}
\end{equation}
, where $G_{ij}=\frac{\partial U_i}{\partial x_j}$ is the mean velocity gradient. The rate of dissipation is evolved via
\begin{equation}
\dot{\epsilon}=-C_0\frac{\epsilon^2}{q^2}-C_sS_{pq}r_{pq}\epsilon-C_{\Omega}\sqrt{\Omega_n\Omega_md_{nm}}\epsilon
\end{equation}
, where $\Omega$ is the mean vorticity vector. The time scale is defined as $\tau=\frac{q^2}{\epsilon}C^vr_{kj}r_{mi}d_{km}$. Based on \citet{iprm2}, the set of coefficients $(C^n, C^v, C_0, C_s, C_{\Omega})$ is fixed at $(2.2, 2.2, 3.67, 3.0, 0.01)$. The slow rotational randomization coefficient $C_r$ is modeled via
\begin{equation}
C_r=\frac{8.5}{\tau}\Omega^*f_{pq}n_pn_q,
\end{equation}
where $\Omega^*=\sqrt{\Omega^*_k \Omega^*_k}$ and $\Omega^*_i=\epsilon_{ipq}r_{qk}d_{kp}$.

The IPRM provides very accurate predictions of single-point statistics in homogeneous turbulence, for a wide variety of mean deformations at disparate values of the normalized strain rate parameter, $\frac{Sk}{\epsilon}$. For instance, the accuracy of the IPRM for the flows considered in this investigation is exhibited in detail in \citet{iprm2,iprm,krr,campos,tf61}. In this light, we utilize the IPRM as our truth model, and, for the purposes of this investigation we discount any uncertainty that might arise from our choice of the model.

In ensemble forecasts used in Numerical Weather Prediction and similar applications, a variety of different techniques have been utilized to generate initial ensembles to reflect the state of uncertainty in the initial state of the system. For instance, the National Centers for Environmental Prediction use breeding of growing vectors to generate samples, utilizing the dominant Local Lyapunov Vectors of the system \citep{zoltan}. Similarly, the European Center for Medium-Range Weather Forecasts, use a collection of singular vectors based on the total energy of the system to represent the uncertainty for initial ensembles \citep{toth}. However, such methods rely on a dynamical systems based analysis of the model evolution equations and an exhaustive identification of local Lyapunov exponents. This is not feasible for the system considered in our study. These approaches select their initial ensembles while considering the Lyapunov exponents so as to maximize separation between pairs of initial conditions. Along the lines of \citet{leith1974}, we generate the initial conditions for this study using maximum entropy distributions, to conserve the measure of uncertainty obligated upon the problem. 

In homogeneous turbulence, the turbulent velocity and pressure fields can be projected into Fourier space via: $
{u'}_i(\mathbf{x},t)=\sum \hat{u}_i(\boldsymbol{\kappa},t)exp(i\boldsymbol{\kappa}\cdot\mathbf{x}), \quad
   {p}^{(r)}(\mathbf{x},t)=\sum{\hat{p}}(\boldsymbol{\kappa},t)exp(i\boldsymbol{\kappa}\cdot\mathbf{x}).$
Thus, the fluctuations are characterized in terms of the wavenumber vector, $\boldsymbol{\kappa}(t)$ and  $\mathbf{\hat{u}}$, $\hat{p}$, the corresponding Fourier amplitudes and pressure coefficients. The Fourier amplitudes and wavenumber evolve due to the action of the applied mean gradient, remaining perpendicular due to the incompressibility condition. Single-point turbulence models do not have access to the states of the individual modes and rely on the singleton statistic of the Reynolds stress tensor to represent this ensemble, $R_{ij}=\sum_{\kappa} \hat{u}^*_i \hat{u}_j$. For any value of the initial Reynolds stress tensor, there exist different possible compositions of the modal ensemble, each corresponding to the same Reynolds stress tensor. This internal structure of a turbulent flow is highly dependent on the history of the flow. Thus, the exact distribution of the wavevectors for a given Reynolds stress is contingent upon the flow history. There is no systematic manner to account for all the possible flow histories and the concomitant set of internal structuring of the flow. In this regard, we utilize maximum entropy distributions \citep{maxent,soize2008} to determine the initial ensembles for the wavevectors in Fourier space. For a continuous random variable $X$, the differential entropy, $h(X)$, can be defined as
\begin{equation}
h(X)=-\int_{X} f_X(x) log(f_X(x))dx,
\end{equation}
where $f_X$ represents the probability density function associated with $X$. The entropy is a measure of the the amount of uncertainty contained in a probability distribution. In keeping with the principle of maximum entropy, from the set of all probability distributions compatible with the level of specification, we choose the ones that maximize differential entropy. For the Fourier amplitudes of the fluctuating velocity field the constraints are given by: $\sum_{\kappa}\hat{u}=0$ and $\sum_{\kappa} \hat{u}^*_i \hat{u}_j=R_{ij}$. With this constraint, the maximum entropy distribution is the multivariate normal distribution, $\hat{u} \backsim \mathcal{N}(0, \Sigma)$. Similarly, once a Fourier amplitude has been determined for a mode, the only constraint on the unit wavevector alignment is that it should lie in the plane perpendicular to the Fourier amplitude to maintain continuity. In this context, the maximum entropy distribution for the alignment of the unit wavenumber vector is the uniform distribution, $\mathbf{e} \backsim \mathcal{U}(0,2\pi)$. This methodology of constructing the initial ensembles is mathematically equivalent to \citet{leith1974}, wherein random perturbations sampling the estimated analysis error covariance are added to the initial conditions. It has been exhibited that using this approach, a relatively small number of integrations is adequate to estimate the statistical properties of the infinite ensemble. Every initial condition for our simulations is constituted of an ensemble of modes, or "particles". The key distinction between different ensembles is the internal structuring of the constituent modes. Each of these ensembles correspond to the same initial componentiality (or the Reynolds stress), but may have different dimensionality, circulicity, etc. Our sensitivity analysis seeks to address the adequacy of the the single-point, second-order characterization of a homogeneous turbulent flow field in terms of the Reynolds stresses only, while ignoring all other higher-order and multi-point moments. An analogous problem occurs in the specification of the near-wall velocity boundary conditions for LES \citep{baggett}. Prior investigations in this problem have shown that detailing correct statistics up to the second order at the boundary is insufficient to obtain the correct core flow. Imposition of just the mean velocities and mean turbulent stresses at the artificial boundaries leads to unsatisfactory agreement, even in simple channel flow simulations. These investigations have concluded that while estimation of moments beyond the second-order is unnecessary, additional information about the structure of the turbulent flow via non-local tensors is imperative to obtain a satisfactory core flow \citep{cabot}.    

\begin{figure}
  \centering
      \includegraphics[height=0.35\textwidth]{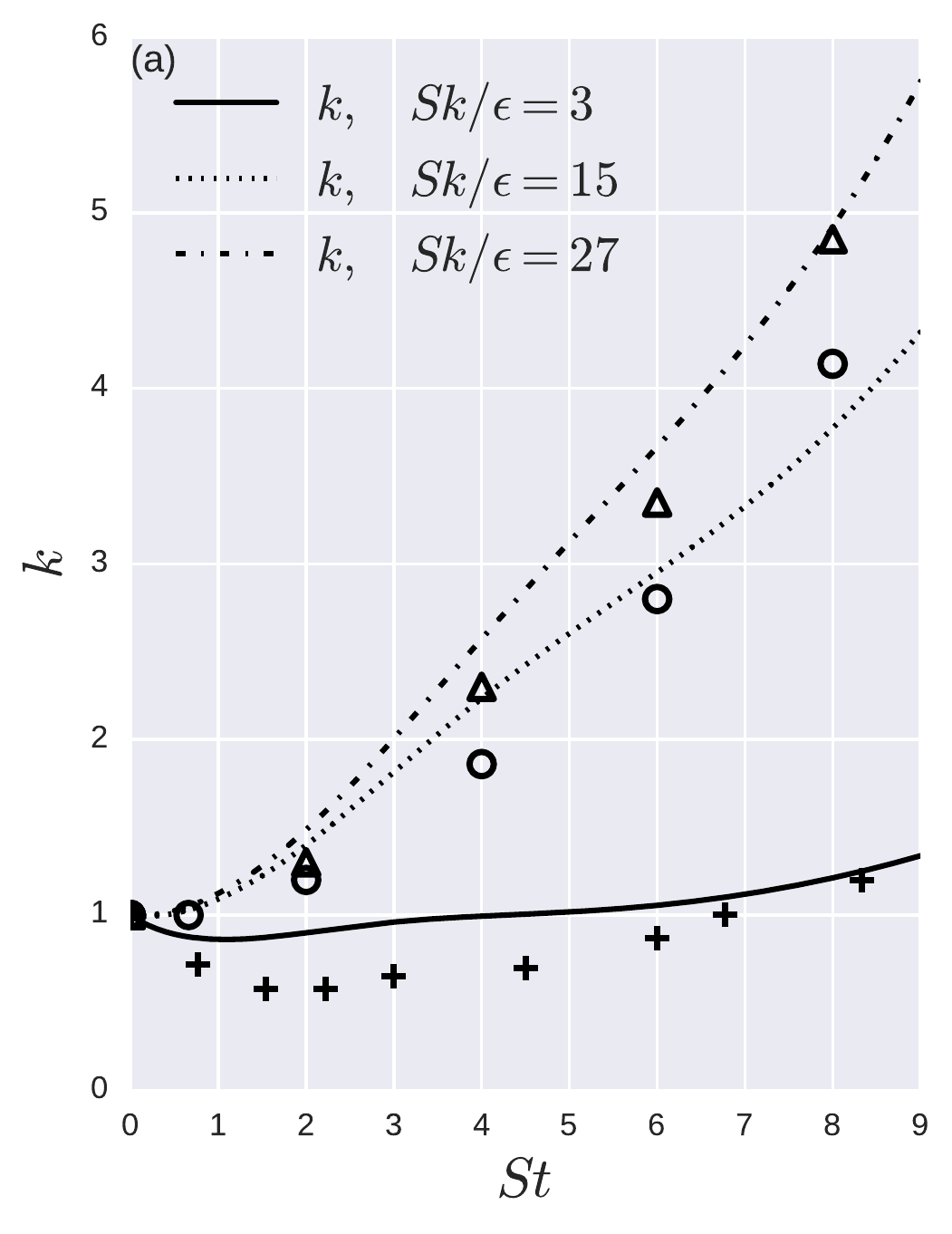}
      \includegraphics[height=0.35\textwidth]{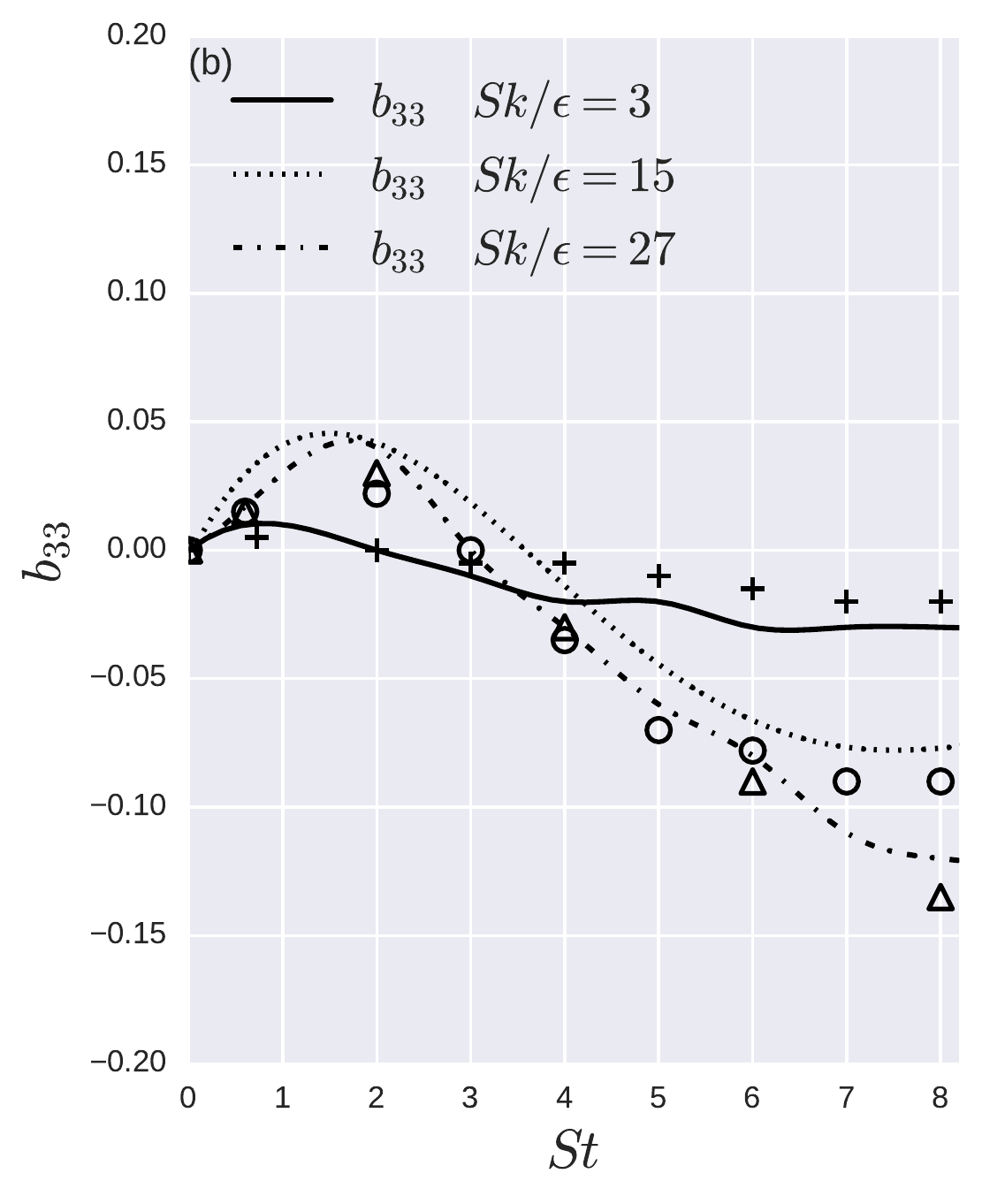}
  \caption{Comparison against the DNS data of \citet{isazancollins}: (a) turbulent kinetic energy evolution, (b) $b_{33}$ evolution. The convention for the symbols is replicated from \cite{isazancollins}, wherein $ +:\frac{Sk}{\epsilon}=3$, $\quad o:\frac{Sk}{\epsilon}=15$ , and, $\quad \triangle:\frac{Sk}{\epsilon}=27$.}
  \label{fig:isazancollins}
\end{figure}

The application of the IPRM model, the methodology for generating the initial conditions and the solution procedure were validated via comparison against numerous high fidelity data sets to ensure that they did not bias the results. Representative results form the comparison against the direct numerical simulations of the HTSF case by \citet{isazancollins} are exhibited in figure \ref{fig:isazancollins}. As can be seen, there is satisfactory agreement between the DNS data set and the mean over the ensembles in our solutions, for both the evolution of the turbulent kinetic energy and the Reynolds stress anisotropies (defined in \cite{isazancollins} as $b_{ij}=\frac{R_{ij}}{2k}-\frac{\delta_{ij}}{3}$). 

For the purposes of this investigation, we restrict ourselves to specific canonical families of mean flows. The planar quadratic flows, that include cases such as plane strain, planar purely sheared flow and purely rotating flow, can be characterized via the ellipticity parameter: $\beta, \left(\frac{W_{ij}W_{ij}}{S_{ij}S_{ij}+W_{ij}W_{ij}}\right)$ \citep{mishra2, mishra3}. The relative strengths of the rate of strain and rotation tensors can be ascertained by non-dimensionalizing their norms by that of the mean velocity gradient tensor. In this regard, we define the derived parameters:
\begin{equation}
a=\sqrt{\frac{1-\beta}{2}} ,\quad
b=\sqrt{\frac{\beta}{2}} .
\end{equation}

\begin{figure}
  \centering
      \includegraphics[width=0.5\textwidth]{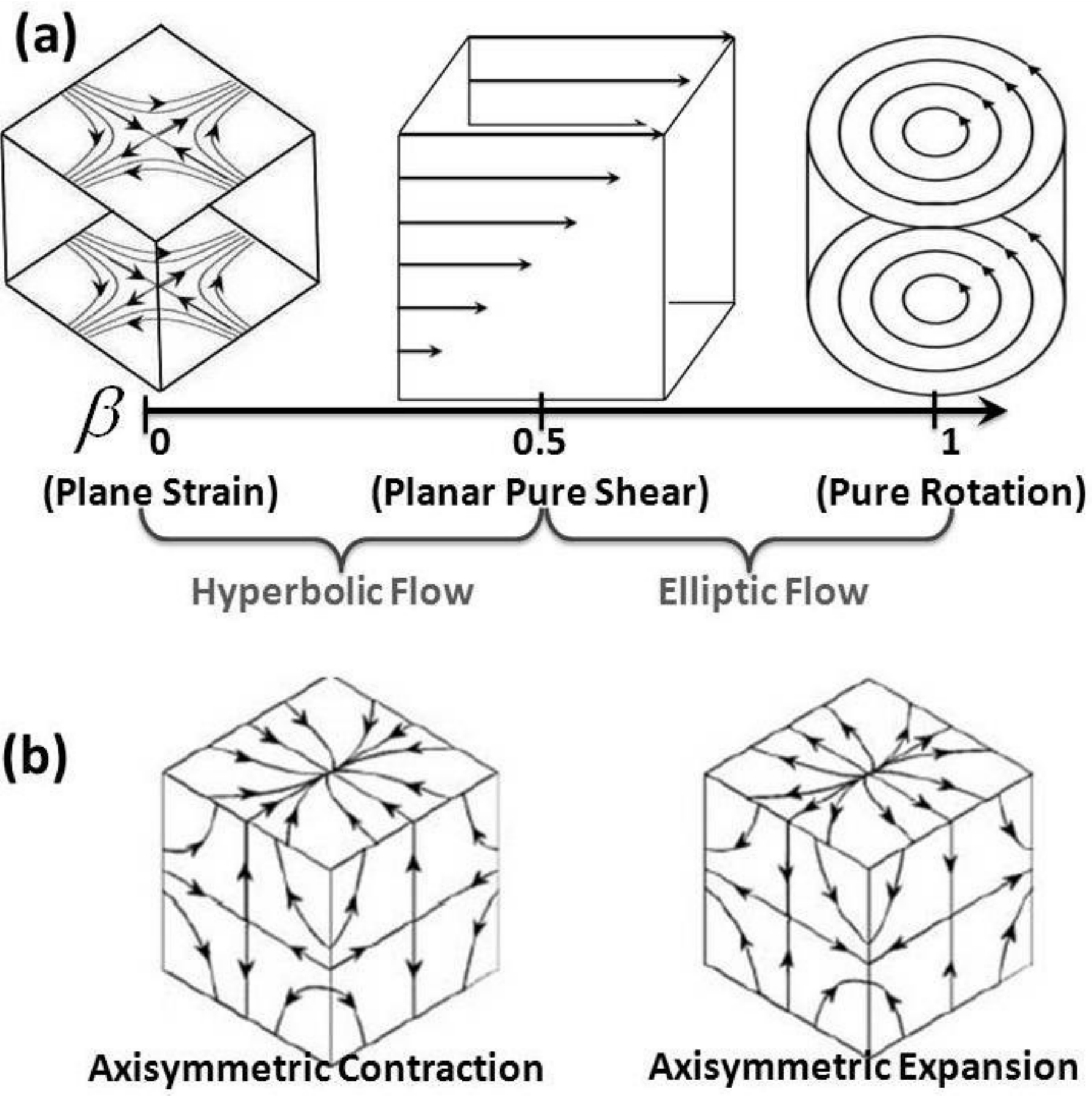}
  \caption{Classification of flow regimes a) Planar quadratic flows, b) Canonical non-planar flows}
\end{figure} 

Here, the derived parameter ``$a$" measures the relative strength of the applied strain and ``$b$", of the applied rotation. The requisite mean flow tensors used in the simulations are given as,
\begin{equation}
\frac{\partial{U}_i}{\partial{x}_j}=\begin{bmatrix}
a& -b& 0\\
b& -a& 0\\
0&  0& 0
\end{bmatrix}.
\end{equation} Thus, for the planar flows, the mean strain is restricted to the $e_1-e_2$ plane and the axis of rotation is aligned along the $e_3$ unit vector. Additionally, we study the non-planar cases of axisymmetric contraction and expansion, where the mean gradient tensors used are given respectively by:
\begin{equation}
\frac{\partial{U}_i}{\partial{x}_j}=\begin{bmatrix}
S& 0& 0\\
0& -\frac{S}{2}& 0\\
0&  0& -\frac{S}{2}
\end{bmatrix}, \quad{}
\frac{\partial{U}_i}{\partial{x}_j}=\begin{bmatrix}
-S& 0& 0\\
0& \frac{S}{2}& 0\\
0&  0& \frac{S}{2}
\end{bmatrix}.
\end{equation}
Using the IPRM as our truth model, we perform a series of Monte Carlo simulations at a range of values of $\frac{\partial U_i}{\partial x_j}$, $R_{ij}$ and $\frac{Sk}{\epsilon}$. For each set of conditions, the simulation samples consist of over 5000 ensembles of over 20,000 individual Fourier modes. While discussing the results, we exhibit prediction intervals for the turbulent kinetic energy and the components of the Reynolds stress anisotropies. For this choice of the applied mean gradient field and the co-ordinate system utilized, the production of the turbulent kinetic energy is given by $\mathcal{P}=a(R_{22}-R_{11})$. Thus, the diagonal components of the Reynolds stress are relatively more important in the evolution of the flow than the off-diagonal components. Accordingly, we exhibit the diagonal components, in the plane of applied shear to highlight the inertial physics, and, perpendicular to the plane of applied shear to highlight the pressure strain interactions. To explicate the observed sensitivity, we utilize contours detailing the relationship between modal stability and modal alignment. These are exhibited in the phase space of the unit wavenumber vector, $\mathbf{e}=\frac{\mathbf{\kappa}}{ \vert \mathbf{\kappa} \vert }$. For hyperbolic flows, the instability coefficient utilized in these contours is defined as $\lambda(t)=log(\frac{k(t)}{k(0)})$, where $k$ is the turbulent kinetic energy of the flow. For elliptic flows, we utilize the Floquet multiplier as a measure of flow instability, described in detail by \citet{hills}.

\section{Results \& Analysis}

Investigations in Numerical Weather Prediction have observed that the predictability of the weather pattern depends on manifold factors. The growth of the initial perturbations in the ensemble may be modest or rapid, dependent upon the local weather patterns \textit{and} the global climate conditions. In the same vein, we assess dependence of the sensitivity of turbulence evolution upon different parameters, specifically, the applied mean gradient, $\frac{\partial U}{\partial x}$, the normalized strain rate parameter, $\frac{Sk}{\epsilon}$ and the initial state of the Reynolds stress tensor, $R_{ij}(t=0)$. To characterize the variability, we utilize ensemble prediction intervals addressing the range of possible evolution of the statistic at specified levels of certitude. 

\begin{figure} 
  \centering
      \includegraphics[width=0.55\textwidth]{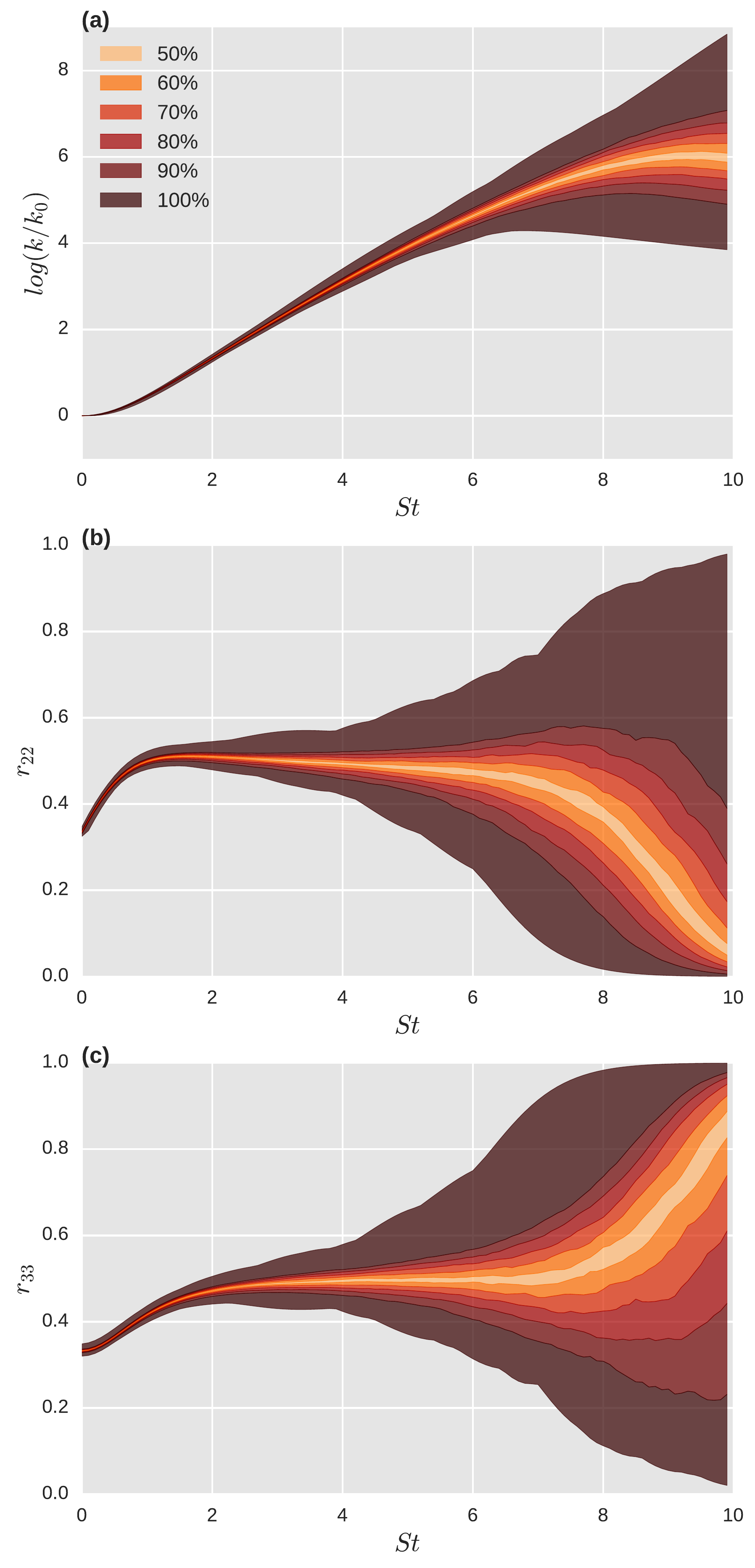}
  \caption{Sample prediction intervals for different levels of certitude for a plane strain mean flow: (a) turbulent kinetic energy, (b) Reynolds stress anisotropy, $r_{22}$, (c) Reynolds stress anisotropy, $r_{33}$. Initially isotropic Reynolds stresses with $\frac{Sk}{\epsilon}=27$.}
\label{fig:pspredints}
\end{figure}

\subsection{Effect of the mean gradient field}

In figure \ref{fig:pspredints}, the prediction intervals for the turbulent kinetic energy, $k$, and the Reynolds stress anisotropies, $r_{ij}=R_{ij}/2k$, are exhibited for a plane strain mean flow ($\beta=0.0$). As can be seen, even after the specification of an isotropic state for the initial Reynolds stress tensor, there is a significant range of possible evolution for the turbulent kinetic energy. Furthermore, after a short stage of moderate growth, this range includes diametric behavior where $k$ can be instantaneously growing for certain ensembles, or, decaying for other choices of modal ensembles. 

This variability in flow evolution is appropriately reflected in the evolution of the Reynolds stress anisotropies as well. In the plane strain mean flow, the $R_{22}$ component has positive production and the inertial effects tend to increase the kinetic energy of this component. If these inertial effects are unchecked, the fluctuations along the $2-$axis would increase unboundedly. Physically, this would lead to a stretching of the Lagrangian fluid particle to very high levels along this direction. To maintain continuity, the pressure strain term transfers this energy to the component perpendicular to the plane of applied shear, $R_{33}$. The production mechanism is completely local and depends only upon the componentiality of the flow. However, the redistribution of turbulent kinetic energy due to the pressure strain term is highly dependent upon the non-local, wavespace information. As the ensembles have the same componentiality, but differ in the internal structuring of the flow, this redistribution of turbulent kinetic energy amongst the components of the Reynolds stress tensor is considerably varied over the ensembles. Thus, there can be initially isotropic ensembles that evolve to a state where the $r_{22}$ component contains almost all the turbulent kinetic energy, or, contrarily, the $r_{22}$ component has an insignificant proportion of the turbulent kinetic energy, as it has been transferred to the $r_{33}$ component due to the pressure strain term. This leads to the broad prediction intervals for $r_{22}$ and $r_{33}$, observed in figures \ref{fig:pspredints} (a) and (b), respectively. 

\begin{figure} 
  \centering
      \includegraphics[width=0.5\textwidth]{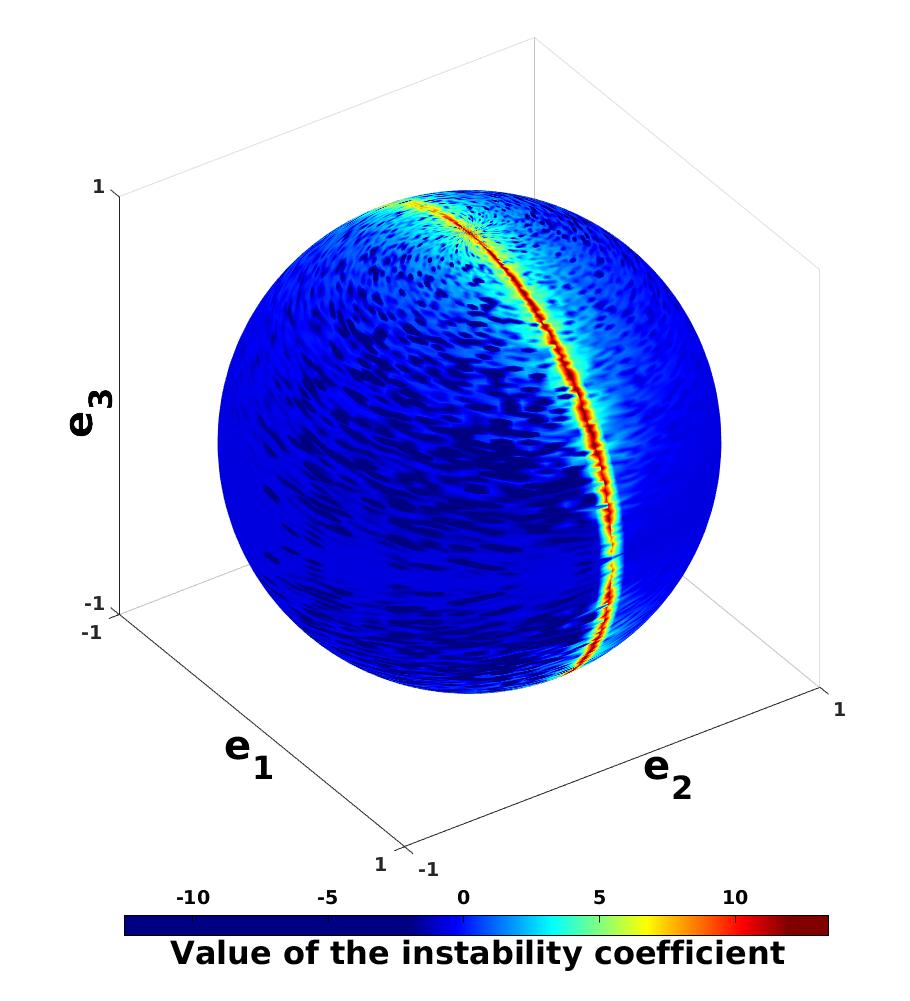}
  \caption{Contours exhibiting the relationship between modal alignment in unit wavenumber space and modal stability, quantified by the instability coefficient, for a plane strain mean flow.}
   \label{fig:psstab}
\end{figure}

This sensitivity can be intuitively understood by considering the modal stability in planar hyperbolic mean flows. Figure \ref{fig:psstab} exhibits the dependence of modal stability upon modal alignment in unit wavenumber space for the case of a plane strain mean flow. As can be seen, \textit{almost all} modes in the flow are stable and the zone of unstable modal alignments is of zero measure. There is an extremely small set of modes, with initial alignments in this zone, that exhibit instability. Furthermore, it is such unstable modes that determine flow statistics. Any variation in the alignments or numbers of these small number of modes would lead to large changes in the evolution of the flow statistics. Consequently, planar hyperbolic flows are very sensitive to any variation in the arrangement of the modes in the ensemble.

\begin{figure}
  \centering
      \includegraphics[width=0.55\textwidth]{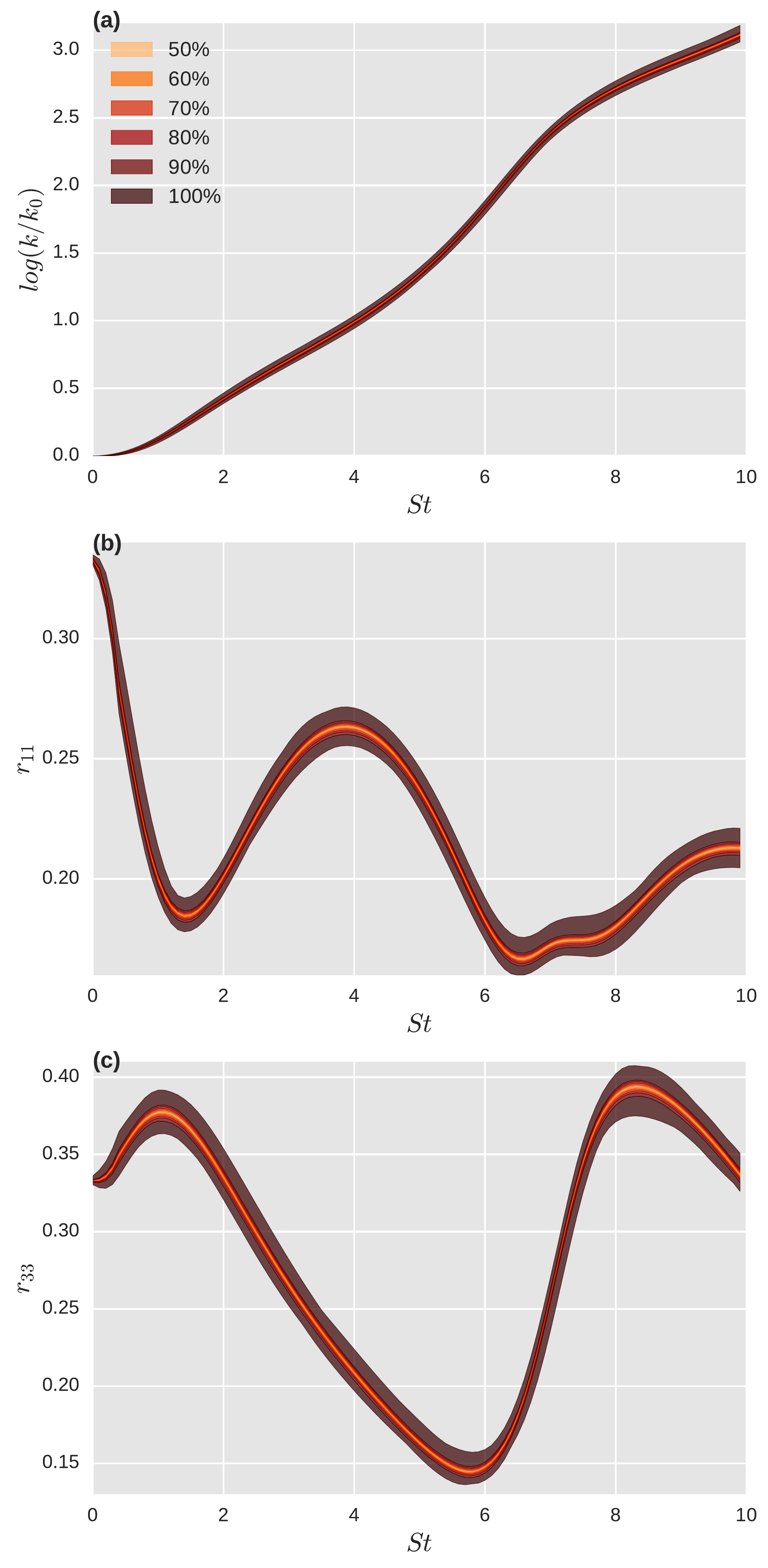}
  \caption{Sample prediction intervals for different levels of certitude for a representative elliptic mean flow: (a) turbulent kinetic energy, (b) Reynolds stress anisotropy, $r_{11}$, (c) Reynolds stress anisotropy, $r_{33}$. Initially isotropic Reynolds stresses with $\frac{Sk}{\epsilon}=27$.}
  \label{fig:ellpredints}
\end{figure} 

In figure \ref{fig:ellpredints}, the prediction intervals for the turbulent kinetic energy, $k$, and the Reynolds stress anisotropies, $r_{ij}=R_{ij}/2k$, are exhibited for a representative elliptic streamline mean flow $(\beta=0.6)$. In spite of the exponential growth due to the elliptic flow instability, the variability in turbulent kinetic evolution is insignificant. Similarly, the Reynolds stress anisotropies do not show appreciable width in their prediction intervals. The flow statistics in elliptic mean flows are relatively robust to the internal structure of the turbulent velocity field.

\begin{figure}
  \centering
      \includegraphics[width=0.5\textwidth]{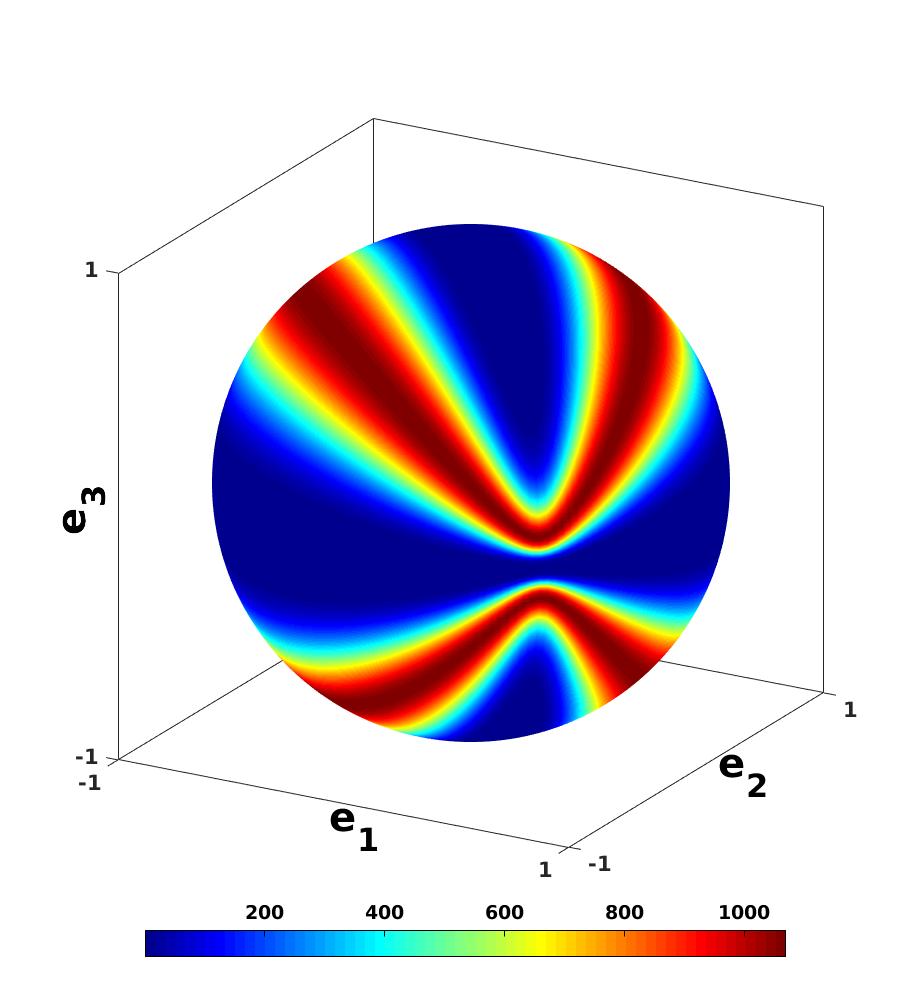}
  \caption{Contours exhibiting the relationship between modal alignment in unit wavenumber space and modal stability, quantified by the Floquet multiplier, for a representative elliptic sheared mean flow.}
  \label{fig:ellstab}
\end{figure}

This robustness can be explicated by considering the dependence of modal stability on modal alignment for elliptic mean flows, exhibited in figure \ref{fig:ellstab}. As has been reported by prior investigations, the elliptic flow instability arises due to parametric resonance, as opposed to the vortex stretching mechanism for hyperbolic flow instability. Due to this mechanism, the zones of instability in unit wavenumber vector space exhibit the characteristic "banded" nature associated with parametric resonance. As the measure of these banded unstable zones is finite and commensurate to the measure of the entire phase space, there is a substantial proportion of modes in the flow that exhibit unstable behavior. Consequently, the evolution of elliptic streamline flows is relatively robust to small, random variations in the arrangement of the modes in the ensemble. 

\begin{figure}
  \centering
      \includegraphics[width=0.55\textwidth]{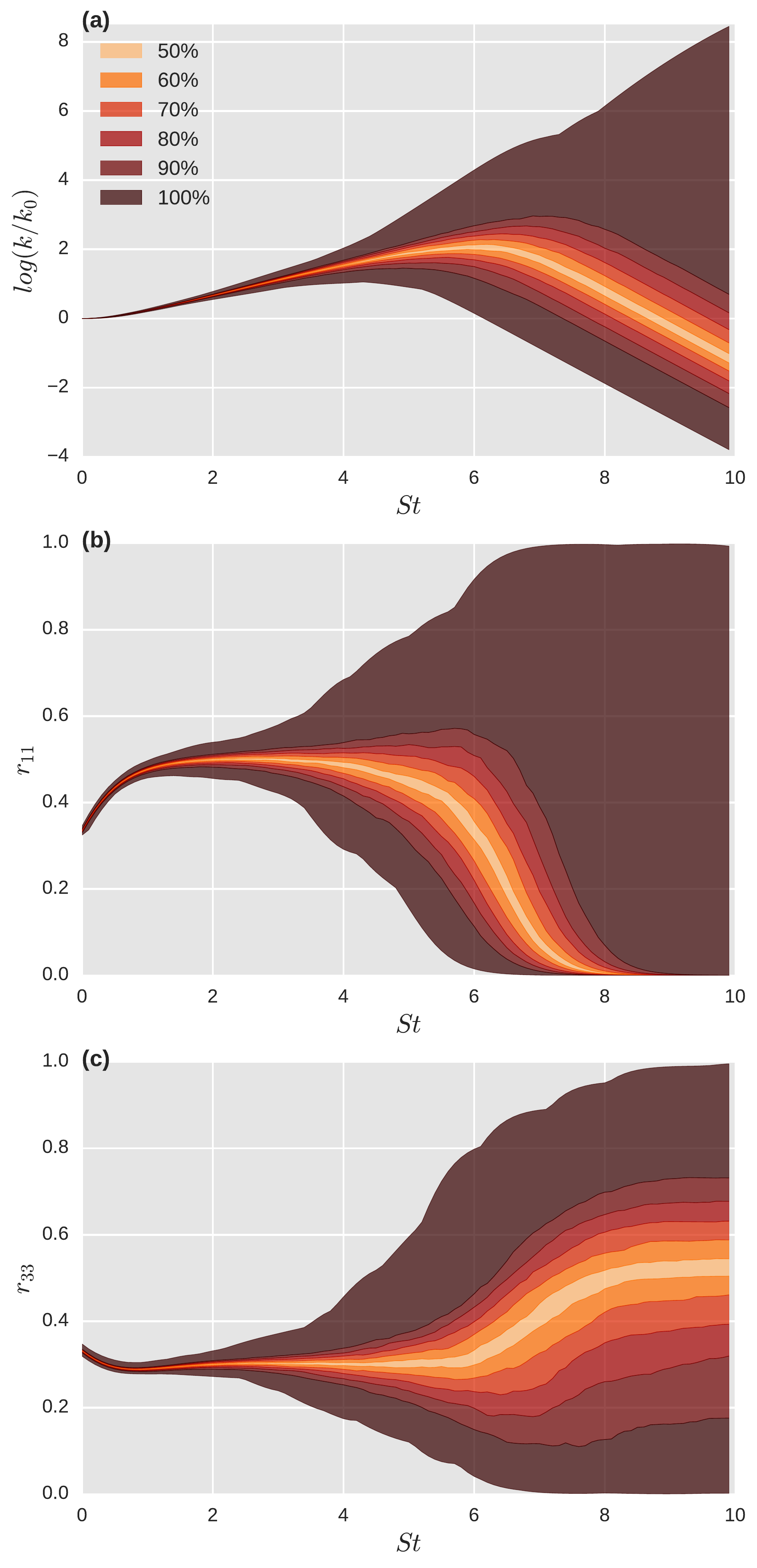}
  \caption{Sample prediction intervals for an axisymmetric expansion mean flow: (a) turbulent kinetic energy, (b) Reynolds stress anisotropy, $r_{11}$, (c) Reynolds stress anisotropy, $r_{33}$. Initially isotropic Reynolds stresses with $\frac{Sk}{\epsilon}=27$.}
  \label{fig:axepredints}
\end{figure}

In figure \ref{fig:axepredints}, the prediction intervals for the turbulent kinetic energy and the Reynolds stress anisotropies are exhibited for a mean flow corresponding to axisymmetric expansion. As can be seen, there is considerable variation in the turbulent kinetic energy evolution for this case. In such a flow, the $R_{11}$ component has positive production while the $R_{22}$ and $R_{33}$ components have negative production. Thus, inertial effects attempt to increase the $R_{11}$ component of the Reynolds stress in an unbounded manner. To maintain continuity, the pressure strain mechanism transfers energy from the $R_{11}$ to the $R_{22}$ and $R_{33}$ components. As this redistribution due to pressure is dependent on non-local information, we observe significant prediction intervals for the evolution of the Reynolds stress anisotropies. Thus, after a short transient, there are ensembles where the pressure strain correlation has transferred most of the turbulent kinetic energy to the $R_{22}$ and $R_{33}$ components. As these Reynolds stress components have negative production, this energy is obliterated, leading to a decay in the turbulent kinetic energy of the flow. Contrarily, there are other modal ensembles where such a redistribution has not occurred appreciably. For these ensembles, the $R_{11}$ component of the Reynolds stress is still energetic and the positive production associated with this component causes the turbulent kinetic energy of the flow to keep growing. 

\begin{figure}
  \centering
      \includegraphics[width=0.5\textwidth]{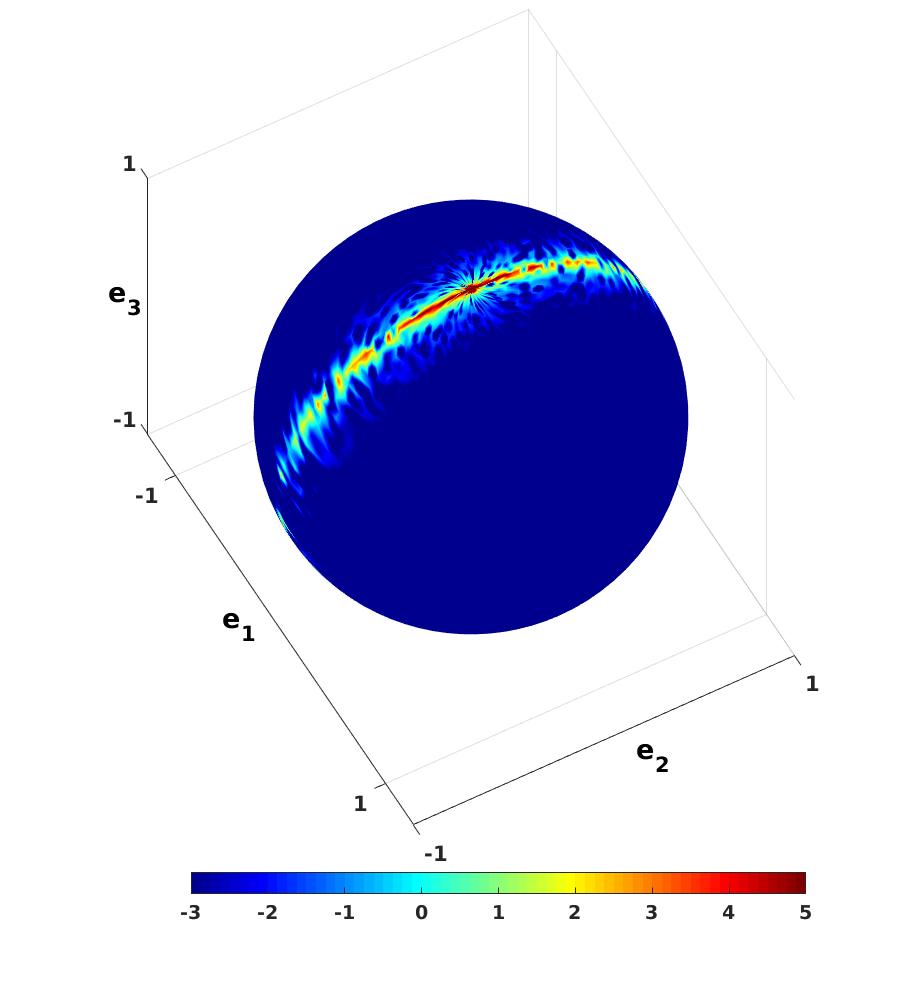}
  \caption{Contours exhibiting the relationship between modal alignment in unit wavenumber space and modal stability for an axisymmetric expansion mean flow.}
   \label{fig:axestab}
\end{figure}

In figure \ref{fig:axestab} the zones of instability are outlined in unit wavenumber vector space for the axisymmetric expansion mean flow. Analogous to the case of a plane strain mean flow, it is observed that \textit{almost all} the modal alignments lead to modal stability. The zone of unstable modal alignments is of zero measure. Thus, the flow statistics are determined by an extremely small set of unstable modes. Any variation in the alignments of these small set of modes would lead to large changes in the evolution of the flow statistics. Consequently, the evolution of the flow is highly sensitive to the internal structure of the modal ensemble. 

\begin{figure}
  \centering
      \includegraphics[width=0.5\textwidth]{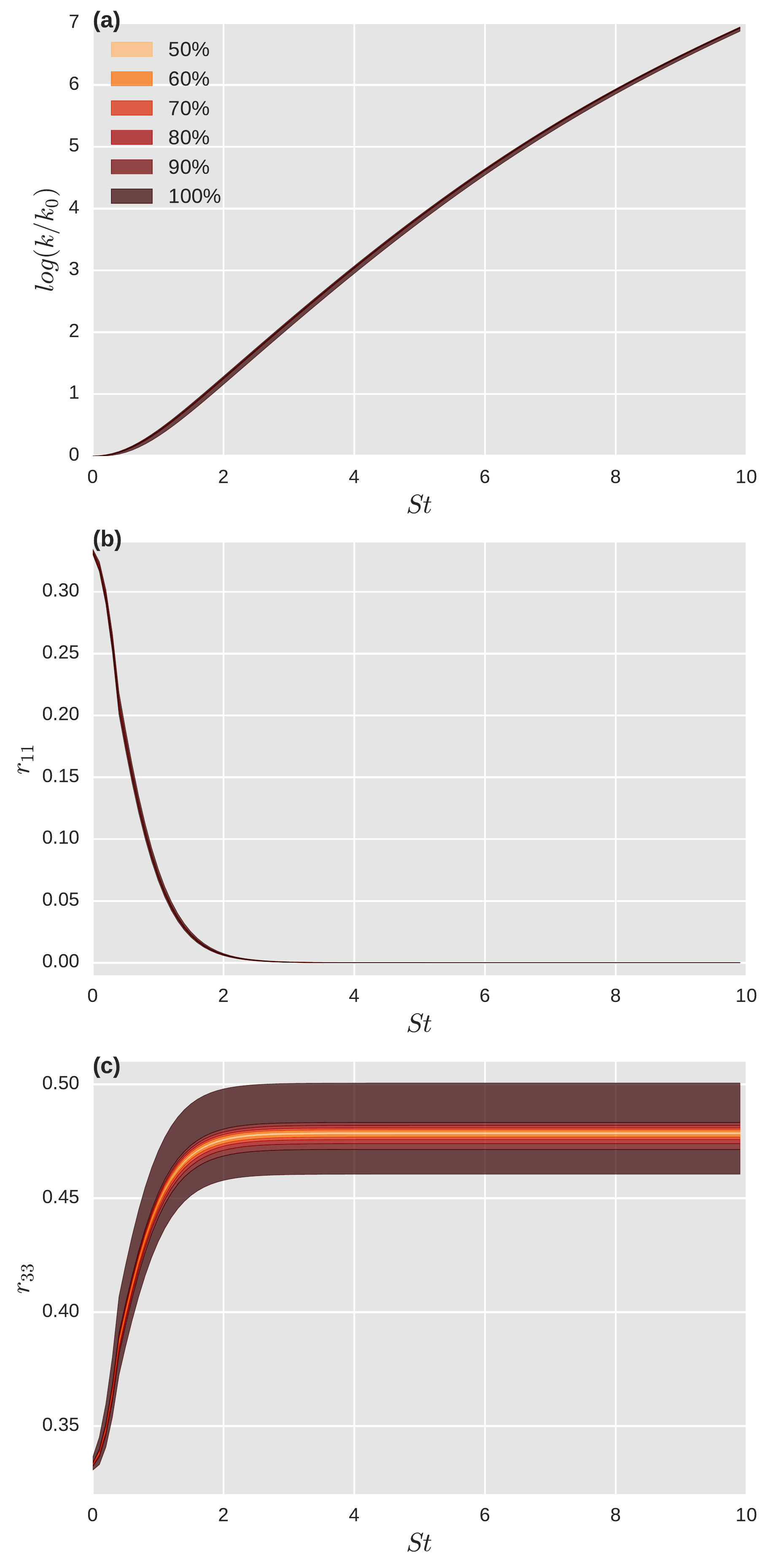}
  \caption{Sample prediction intervals for an axisymmetric contraction mean flow: (a) turbulent kinetic energy, (b) Reynolds stress anisotropy, $r_{11}$, (c) Reynolds stress anisotropy, $r_{33}$. Initially isotropic Reynolds stresses with $\frac{Sk}{\epsilon}=27$.}
  \label{fig:axcpredints}
\end{figure}

In figure \ref{fig:axcpredints}, the prediction intervals for the turbulent kinetic energy and the Reynolds stress anisotropies are exhibited for a mean flow corresponding to axisymmetric contraction. The evolution of the turbulent kinetic energy is relatively robust and unaffected by the internal structuring of the turbulent flow field. Similarly, the Reynolds stress anisotropies do not show significant prediction intervals in their evolution either.

\begin{figure}
  \centering
      \includegraphics[width=0.5\textwidth]{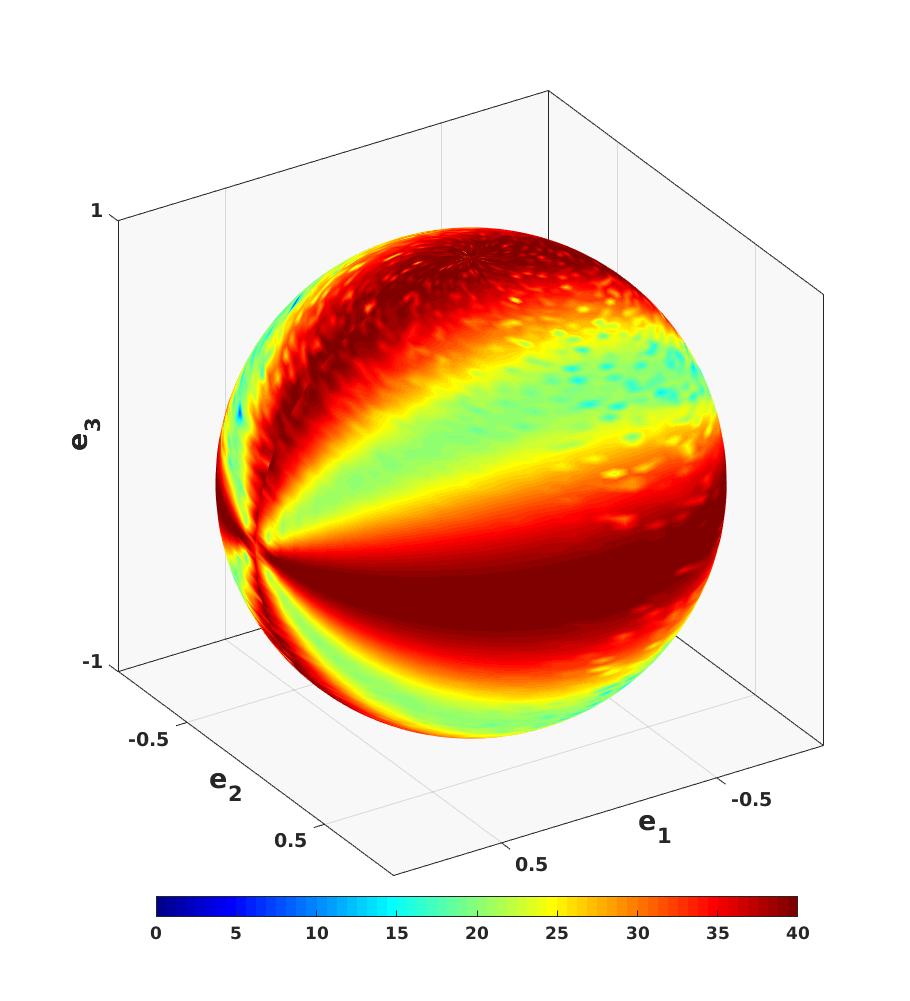}
  \caption{Contours exhibiting the relationship between modal alignment in unit wavenumber space and modal stability for an axisymmetric contraction mean flow.}
  \label{fig:axcstab}
\end{figure}

The zones of unstable initial alignments for this mean flow are outlined in figure \ref{fig:axcstab}. As can be observed, the unstable zones are of finite measure and thus, there is a substantial ratio of modes in the flow that exhibit unstable behavior. Accordingly, the evolution of a flow under axisymmetric contraction is robust to variations in the arrangement of the modes in the ensemble. 

Thus, the sensitivity of flow evolution upon the internal structure of the turbulent flow field is highly dependent upon the mean flow. For mean flows like the planar hyperbolic streamline flow or the three-dimensional case of axisymmetric expansion, the flow evolution is extremely dependent upon the initial arrangement of the modes in the ensemble. However, for mean flows like the planar elliptic streamline flow or the three-dimensional case of axisymmetric contraction, the flow evolution is relatively unaffected by the arrangement of modes in the ensemble. 

The key physics underlying this variation is the non-local effects of the pressure strain correlation term. The inertial production mechanism is completely local and is unaffected by the alignments of modes in the initial ensemble, or, the initial distribution of turbulent kinetic energy amongst these modes. The pressure strain correlation term redistributes the energy between the different components of the modal velocity to ensure continuity, and, is highly dependent upon the modal alignment. This redistribution leads to a "structuring effect" in the modal ensemble wherein the modal stability becomes dependent upon the modal alignment. This divides the unit wavevector space into zones of stable alignments and unstable alignments. Herein, the zones of instability are relevant as the unstable modes in the flow tend to determine flow statistics. When the zones of unstable alignments are of very small measure, very few modes in the ensemble may be unstable and the flow evolution is highly dependent upon initial modal alignment. If the zones of unstable alignments are of sizable measure, a significant proportion of the modes in the ensemble are unstable and the flow evolution is relatively unaffected by the internal structuring of the flow field.

\subsection{Effect of the normalized strain rate}

In a homogeneous turbulent flow, the turbulent transport term is inactive due to spatial statistical homogeneity. The production term affects all modes similarly, irrespective of their alignments. While the dissipation term acts as a sink for turbulent kinetic energy, it affects modes dependent only upon their magnitude. Thus, the slow and the rapid components of the pressure strain correlation are key towards engendering the sensitivity of flow evolution.  

The Poisson equation for pressure in a turbulent flow is given by:
\begin{equation}
\frac{1}{\rho}\nabla^2 p=\underbrace{-2\frac{\partial U_i}{\partial x_j}\frac{\partial u_j}{\partial x_i}}_{\text{Linear/Rapid}}-\underbrace{\frac{\partial^2}{\partial x_i \partial x_j}(u_iu_j-\left\langle u_iu_j \right\rangle)}_{\text{Non-linear/slow}},
\end{equation}
wherein the lowercase symbols represent the fluctuating fields in physical space, the mean fields are denoted by uppercase symbols and $\left\langle . \right\rangle$ represents the ensemble averaging operator. With regard to the action of pressure, by precedent fluctuating pressure is divided into two components, viz. rapid and slow pressure. Mathematically, the terms rapid and slow refer to the components of pressure arising from the linear and non-linear parts of the source term in the Poisson equation for pressure. The slow component acts to conserve the incompressibility of the
velocity field generated by the nonlinear interactions among velocity fluctuations. It is
the function of rapid pressure to impose the divergence free condition on the fluctuating
velocity field produced by linear interactions between the mean and fluctuating fields. The pressure strain correlation is denoted by:
\begin{equation}
\pi_{ij}=\left\langle \frac{p}{\rho}(\frac{\partial u_i}{\partial x_j}+\frac{\partial u_j}{\partial x_i})\right\rangle,
\end{equation}
 In accordance to the decomposition of pressure into a rapid and slow component, the pressure strain correlation can also be split into two contributions: the rapid pressure strain correlation, $\pi^{(r)}_{ij}$, and the slow pressure strain correlation, $\pi^{(s)}_{ij}$. The normalized strain rate parameter provides an indication of the relative strengths of the linear and the non-linear physics. At the rapid distortion limit, $\frac{Sk}{\epsilon}\rightarrow \infty$ and the non-linear pressure strain effects can be ignored. Contrarily, for decaying turbulence,  $\frac{Sk}{\epsilon}\rightarrow 0$ and the linear pressure strain effects can be ignored. Thus, analysis of the dependence of uncertainty on $\frac{Sk}{\epsilon}$ would indicate the relative importance of the non-linear and linear physics to this sensitivity. 
 
\begin{figure}
  \centering
      \includegraphics[width=1\textwidth]{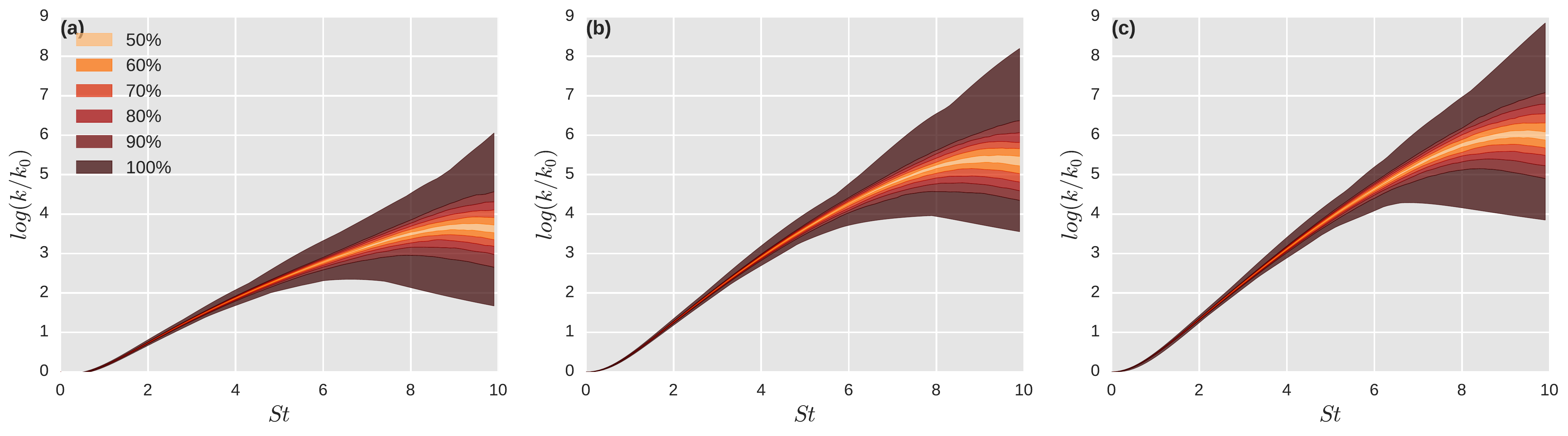}
  \caption{Variation in Prediction intervals for the evolution of turbulent kinetic energy for a plane strain mean flow with (a) $\frac{Sk_0}{\epsilon_0}=3$, (b) $\frac{Sk_0}{\epsilon_0}=15$, (c) $\frac{Sk_0}{\epsilon_0}=27$}
  \label{fig:3bps}
\end{figure}

 As can be seen in figures \ref{fig:3bps}, with increment in the value of $\frac{Sk}{\epsilon}$, the width of the prediction intervals increases for a plane strain mean flow. Similarly, for a flow under axisymmetric expansion, an increment in $\frac{Sk}{\epsilon}$ leads to an increase in the width of the prediction intervals. The effect of the non-linear physics is equivalent to the diffusion of the turbulent kinetic energy towards a uniform, isotropic distribution. This diffusion leads to an accretion in the zones of unstable modal alignments, causing them to obtain non-zero, finite measure. This engenders a reduced sensitivity to the internal structuring of the turbulent flow for hyperbolic flows and a resultant reduction in the prediction intervals.
 
\begin{figure}
  \centering
      \includegraphics[width=0.6\textwidth]{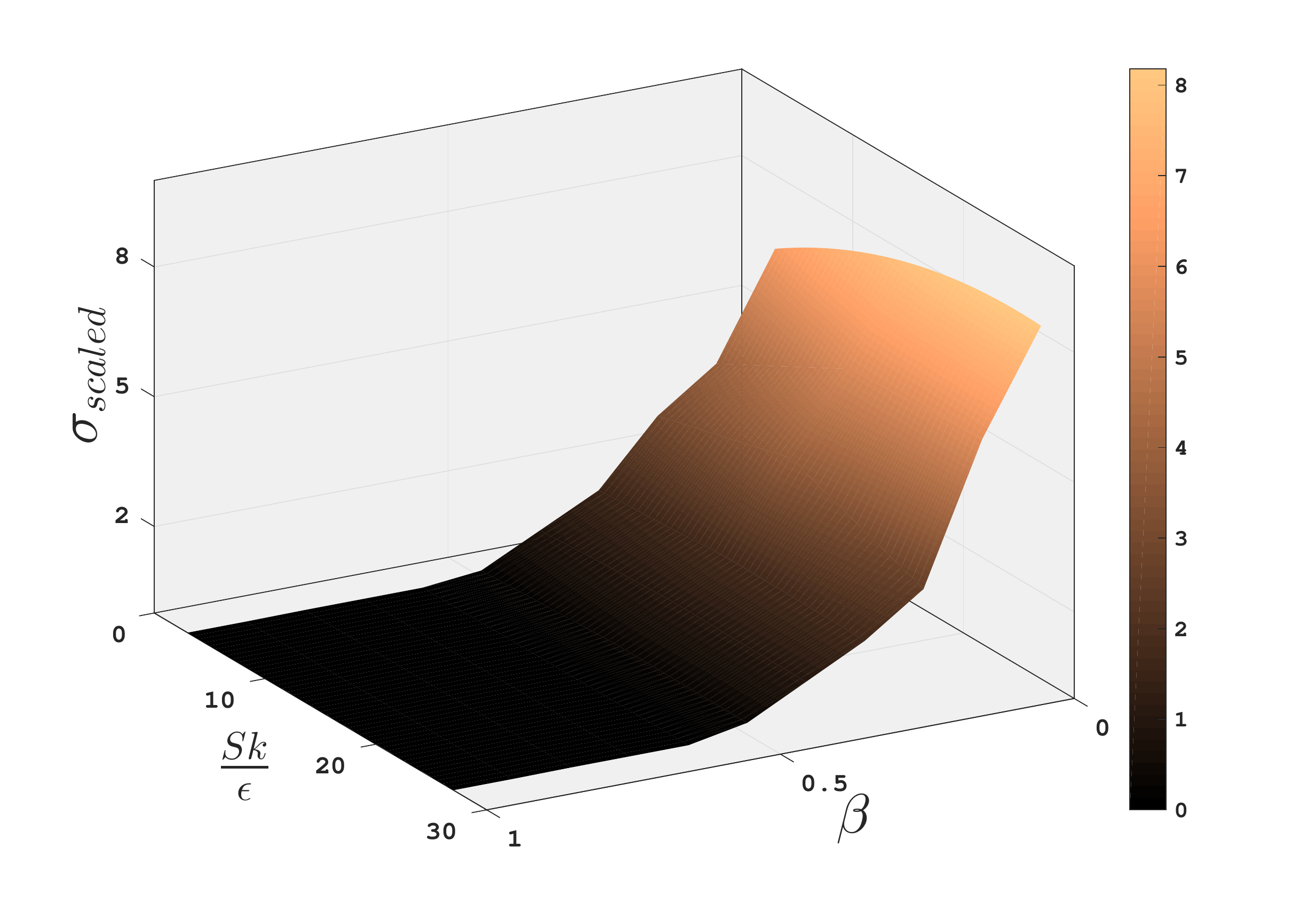}
  \caption{Surface contours delineating the dependence of the uncertainty, quantified by $\sigma_{scaled}(St=10)$, on the applied mean gradient and the strain rate parameter.}
  \label{fig:surface}
\end{figure}

To characterize the dependence of the \textit{uncertainty} on $\frac{Sk}{\epsilon}$, independent of the growth of the instability, we define the measure $\sigma_{scaled} (St)=\frac{k_{max}(St)-k_{min}(St)}{k_{mean}(St)}$, where the turbulent kinetic energies correspond to the different sample ensembles, having the same initially isotropic Reynolds stress tensor but differing in the internal structuring of the turbulent velocity field. Scaling the range of turbulent kinetic evolution amongst the ensembles by the mean over the ensembles ensures that the effect of the rate of turbulent kinetic energy growth for the flow does not unduly bias the measure. As can be seen in Figure \ref{fig:surface}, the uncertainty in predictions reduces with decrease in the value of $\frac{Sk}{\epsilon}$. However, the dependence of the variability on the normalized strain rate is much lower than its dependence upon the applied mean velocity gradient. 
 
\begin{figure}
  \centering
      \includegraphics[width=1\textwidth]{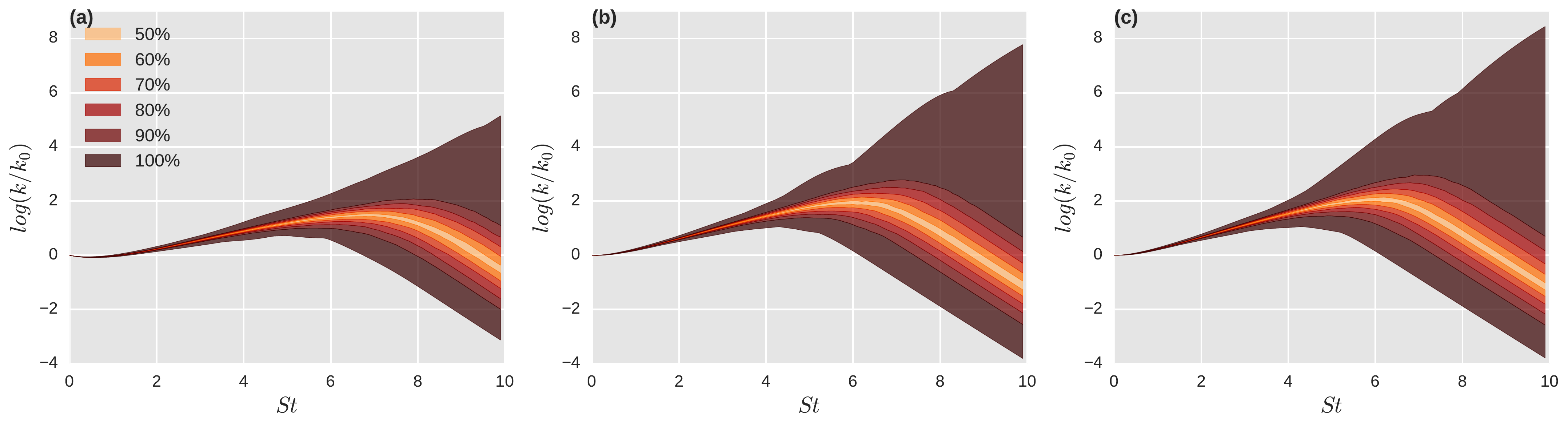}
  \caption{Variation in Prediction intervals for the evolution of turbulent kinetic energy for an axisymmetric expansion mean flow with (a) $\frac{Sk_0}{\epsilon_0}=3$, (b) $\frac{Sk_0}{\epsilon_0}=15$, (c) $\frac{Sk_0}{\epsilon_0}=27$}
  \label{fig:3baxe}
\end{figure}

\subsection{Effect of the initial Reynolds stress tensor}

\begin{figure}
  \centering
      \includegraphics[width=1\textwidth]{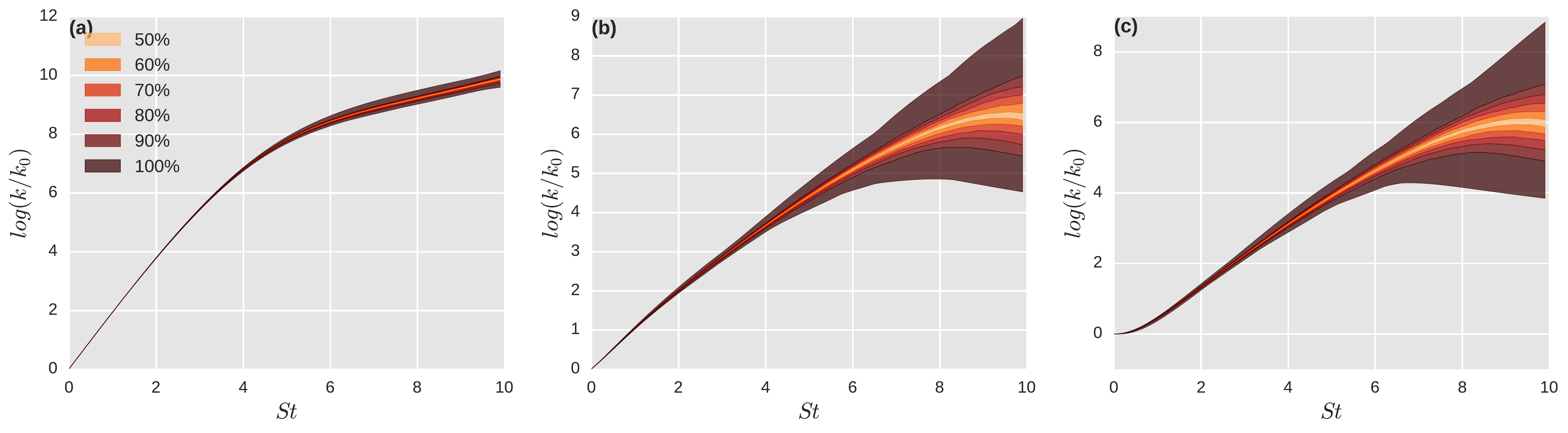}
  \caption{Variation in Prediction intervals for the evolution of turbulent kinetic energy for an plane strain mean flow with (a) an initially 1C state, (b) an initially 2C state, (c) an initially 3C state.}
  \label{fig:3c}
\end{figure}

Additionally, the magnitude and evolution of the prediction intervals depend on the initial state of the Reynolds stress tensor. This is exhibited in figure \ref{fig:3c}, for the case of a plane strain mean flow. The width of the prediction intervals for a turbulent flow with an initial componentiality corresponding to that of the $2C$ limit is lesser than that of the $3C$ limit. In the same trend, the prediction intervals for the flow  with an initial componentiality corresponding to that of the $1C$ limit are relatively negligible. 

\section{Summary \& Conclusions}
In this article, we investigated the sensitivity of flow evolution upon the internal structure of the turbulent flow field. Specifically, we focus upon the fact that the specification of the macro state, via the Reynolds stress tensor, does not lead to a complete and unique specification of the micro state, the internal structuring of the turbulent flow. This lack of adequate description leads to a variability in the evolution of the turbulent flow, causing a degree of uncertainty in the predictive fidelity of RANS models. It was exhibited that the ensuing prediction intervals are significant in magnitude and exhibit temporal growth. Their evolution and their dependence on flow parameters was analyzed and explicated. This identification provides for an a priori, physics based approach toward determining eigenvalue perturbations for uncertainty quantification. In addition to its importance to the formulation of single-point RANS turbulence models, these results help guide the development of improved discrepancy models. \\ \\
\textbf{Acknowledgement}: This research was supported by the Defense Advanced Research Projects Agency under the Enabling Quantification of Uncertainty in Physical Systems (\emph{EQUiPS}) project (technical monitor: Dr Fariba Fahroo).

\bibliography{PRFEnsembleForecasting}

\begin{thebibliography}{39}%
\makeatletter
\providecommand \@ifxundefined [1]{%
 \@ifx{#1\undefined}
}%
\providecommand \@ifnum [1]{%
 \ifnum #1\expandafter \@firstoftwo
 \else \expandafter \@secondoftwo
 \fi
}%
\providecommand \@ifx [1]{%
 \ifx #1\expandafter \@firstoftwo
 \else \expandafter \@secondoftwo
 \fi
}%
\providecommand \natexlab [1]{#1}%
\providecommand \enquote  [1]{``#1''}%
\providecommand \bibnamefont  [1]{#1}%
\providecommand \bibfnamefont [1]{#1}%
\providecommand \citenamefont [1]{#1}%
\providecommand \href@noop [0]{\@secondoftwo}%
\providecommand \href [0]{\begingroup \@sanitize@url \@href}%
\providecommand \@href[1]{\@@startlink{#1}\@@href}%
\providecommand \@@href[1]{\endgroup#1\@@endlink}%
\providecommand \@sanitize@url [0]{\catcode `\\12\catcode `\$12\catcode
  `\&12\catcode `\#12\catcode `\^12\catcode `\_12\catcode `\%12\relax}%
\providecommand \@@startlink[1]{}%
\providecommand \@@endlink[0]{}%
\providecommand \url  [0]{\begingroup\@sanitize@url \@url }%
\providecommand \@url [1]{\endgroup\@href {#1}{\urlprefix }}%
\providecommand \urlprefix  [0]{URL }%
\providecommand \Eprint [0]{\href }%
\providecommand \doibase [0]{http://dx.doi.org/}%
\providecommand \selectlanguage [0]{\@gobble}%
\providecommand \bibinfo  [0]{\@secondoftwo}%
\providecommand \bibfield  [0]{\@secondoftwo}%
\providecommand \translation [1]{[#1]}%
\providecommand \BibitemOpen [0]{}%
\providecommand \bibitemStop [0]{}%
\providecommand \bibitemNoStop [0]{.\EOS\space}%
\providecommand \EOS [0]{\spacefactor3000\relax}%
\providecommand \BibitemShut  [1]{\csname bibitem#1\endcsname}%
\let\auto@bib@innerbib\@empty
\bibitem [{\citenamefont {Ling}\ and\ \citenamefont {Templeton}(2015)}]{ling1}%
  \BibitemOpen
  \bibfield  {author} {\bibinfo {author} {\bibfnamefont {J.}~\bibnamefont
  {Ling}}\ and\ \bibinfo {author} {\bibfnamefont {J.}~\bibnamefont
  {Templeton}},\ }\bibfield  {title} {\enquote {\bibinfo {title} {Evaluation of
  machine learning algorithms for prediction of regions of high reynolds
  averaged navier stokes uncertainty},}\ }\href@noop {} {\bibfield  {journal}
  {\bibinfo  {journal} {Physics of Fluids (1994-present)}\ }\textbf {\bibinfo
  {volume} {27}},\ \bibinfo {pages} {085103} (\bibinfo {year}
  {2015})}\BibitemShut {NoStop}%
\bibitem [{\citenamefont {Edeling}\ \emph
  {et~al.}(2014{\natexlab{a}})\citenamefont {Edeling}, \citenamefont
  {Cinnella}, \citenamefont {Dwight},\ and\ \citenamefont {Bijl}}]{edelinga}%
  \BibitemOpen
  \bibfield  {author} {\bibinfo {author} {\bibfnamefont {W.~N.}\ \bibnamefont
  {Edeling}}, \bibinfo {author} {\bibfnamefont {P.}~\bibnamefont {Cinnella}},
  \bibinfo {author} {\bibfnamefont {R.~P.}\ \bibnamefont {Dwight}}, \ and\
  \bibinfo {author} {\bibfnamefont {H.}~\bibnamefont {Bijl}},\ }\bibfield
  {title} {\enquote {\bibinfo {title} {Bayesian estimates of parameter
  variability in the k--$\varepsilon$ turbulence model},}\ }\href@noop {}
  {\bibfield  {journal} {\bibinfo  {journal} {Journal of Computational
  Physics}\ }\textbf {\bibinfo {volume} {258}},\ \bibinfo {pages} {73--94}
  (\bibinfo {year} {2014}{\natexlab{a}})}\BibitemShut {NoStop}%
\bibitem [{\citenamefont {Margheri}\ \emph {et~al.}(2014)\citenamefont
  {Margheri}, \citenamefont {Meldi}, \citenamefont {Salvetti},\ and\
  \citenamefont {Sagaut}}]{margheri}%
  \BibitemOpen
  \bibfield  {author} {\bibinfo {author} {\bibfnamefont {L.}~\bibnamefont
  {Margheri}}, \bibinfo {author} {\bibfnamefont {M.}~\bibnamefont {Meldi}},
  \bibinfo {author} {\bibfnamefont {M.~V.}\ \bibnamefont {Salvetti}}, \ and\
  \bibinfo {author} {\bibfnamefont {P.}~\bibnamefont {Sagaut}},\ }\bibfield
  {title} {\enquote {\bibinfo {title} {Epistemic uncertainties in rans model
  free coefficients},}\ }\href@noop {} {\bibfield  {journal} {\bibinfo
  {journal} {Computers \& Fluids}\ }\textbf {\bibinfo {volume} {102}},\
  \bibinfo {pages} {315--335} (\bibinfo {year} {2014})}\BibitemShut {NoStop}%
\bibitem [{\citenamefont {Edeling}\ \emph
  {et~al.}(2014{\natexlab{b}})\citenamefont {Edeling}, \citenamefont
  {Cinnella},\ and\ \citenamefont {Dwight}}]{edelingb}%
  \BibitemOpen
  \bibfield  {author} {\bibinfo {author} {\bibfnamefont {W.~N.}\ \bibnamefont
  {Edeling}}, \bibinfo {author} {\bibfnamefont {P.}~\bibnamefont {Cinnella}}, \
  and\ \bibinfo {author} {\bibfnamefont {R.~P.}\ \bibnamefont {Dwight}},\
  }\bibfield  {title} {\enquote {\bibinfo {title} {Predictive rans simulations
  via bayesian model-scenario averaging},}\ }\href@noop {} {\bibfield
  {journal} {\bibinfo  {journal} {Journal of Computational Physics}\ }\textbf
  {\bibinfo {volume} {275}},\ \bibinfo {pages} {65--91} (\bibinfo {year}
  {2014}{\natexlab{b}})}\BibitemShut {NoStop}%
\bibitem [{\citenamefont {Jones}\ and\ \citenamefont
  {Launder}(1972)}]{joneslaunder}%
  \BibitemOpen
  \bibfield  {author} {\bibinfo {author} {\bibfnamefont {W.~P.}\ \bibnamefont
  {Jones}}\ and\ \bibinfo {author} {\bibfnamefont {B.~E.}\ \bibnamefont
  {Launder}},\ }\bibfield  {title} {\enquote {\bibinfo {title} {The prediction
  of laminarization with a two-equation model of turbulence},}\ }\href@noop {}
  {\bibfield  {journal} {\bibinfo  {journal} {International journal of heat and
  mass transfer}\ }\textbf {\bibinfo {volume} {15}},\ \bibinfo {pages}
  {301--314} (\bibinfo {year} {1972})}\BibitemShut {NoStop}%
\bibitem [{\citenamefont {Mishra}\ and\ \citenamefont
  {Girimaji}(2010)}]{mishra1}%
  \BibitemOpen
  \bibfield  {author} {\bibinfo {author} {\bibfnamefont {A.~A.}\ \bibnamefont
  {Mishra}}\ and\ \bibinfo {author} {\bibfnamefont {S.~S.}\ \bibnamefont
  {Girimaji}},\ }\bibfield  {title} {\enquote {\bibinfo {title}
  {Pressure--strain correlation modeling: towards achieving consistency with
  rapid distortion theory},}\ }\href@noop {} {\bibfield  {journal} {\bibinfo
  {journal} {Flow, turbulence and combustion}\ }\textbf {\bibinfo {volume}
  {85}},\ \bibinfo {pages} {593--619} (\bibinfo {year} {2010})}\BibitemShut
  {NoStop}%
\bibitem [{\citenamefont {Kassinos}\ and\ \citenamefont
  {Reynolds}(1994)}]{tf61}%
  \BibitemOpen
  \bibfield  {author} {\bibinfo {author} {\bibfnamefont {S.~C.}\ \bibnamefont
  {Kassinos}}\ and\ \bibinfo {author} {\bibfnamefont {W.~C.}\ \bibnamefont
  {Reynolds}},\ }\href@noop {} {\enquote {\bibinfo {title} {A structure-based
  model for the rapid distortion of homogeneous turbulence. report tf-61,
  thermosciences division, department of mechanical engineering},}\ } (\bibinfo
  {year} {1994})\BibitemShut {NoStop}%
\bibitem [{\citenamefont {Cambon}\ and\ \citenamefont
  {Rubinstein}(2006)}]{cambon2006}%
  \BibitemOpen
  \bibfield  {author} {\bibinfo {author} {\bibfnamefont {C.}~\bibnamefont
  {Cambon}}\ and\ \bibinfo {author} {\bibfnamefont {R.}~\bibnamefont
  {Rubinstein}},\ }\bibfield  {title} {\enquote {\bibinfo {title} {Anisotropic
  developments for homogeneous shear flows},}\ }\href@noop {} {\bibfield
  {journal} {\bibinfo  {journal} {Physics of Fluids (1994-present)}\ }\textbf
  {\bibinfo {volume} {18}},\ \bibinfo {pages} {085106} (\bibinfo {year}
  {2006})}\BibitemShut {NoStop}%
\bibitem [{\citenamefont {Kraichnan}(1959)}]{kraichnan}%
  \BibitemOpen
  \bibfield  {author} {\bibinfo {author} {\bibfnamefont {R.~H.}\ \bibnamefont
  {Kraichnan}},\ }\bibfield  {title} {\enquote {\bibinfo {title} {The structure
  of isotropic turbulence at very high reynolds numbers},}\ }\href@noop {}
  {\bibfield  {journal} {\bibinfo  {journal} {Journal of Fluid Mechanics}\
  }\textbf {\bibinfo {volume} {5}},\ \bibinfo {pages} {497--543} (\bibinfo
  {year} {1959})}\BibitemShut {NoStop}%
\bibitem [{\citenamefont {Rubinstein}\ and\ \citenamefont
  {Girimaji}(2006)}]{rng}%
  \BibitemOpen
  \bibfield  {author} {\bibinfo {author} {\bibfnamefont {R.}~\bibnamefont
  {Rubinstein}}\ and\ \bibinfo {author} {\bibfnamefont {S.~S.}\ \bibnamefont
  {Girimaji}},\ }\bibfield  {title} {\enquote {\bibinfo {title} {Second moment
  closure near the two-component limit},}\ }\href@noop {} {\bibfield  {journal}
  {\bibinfo  {journal} {Journal of Fluid Mechanics}\ }\textbf {\bibinfo
  {volume} {548}},\ \bibinfo {pages} {197--206} (\bibinfo {year}
  {2006})}\BibitemShut {NoStop}%
\bibitem [{\citenamefont {Hadamard}(1902)}]{Hadamard}%
  \BibitemOpen
  \bibfield  {author} {\bibinfo {author} {\bibfnamefont {J.}~\bibnamefont
  {Hadamard}},\ }\bibfield  {title} {\enquote {\bibinfo {title} {Sur les
  probl{\`e}mes aux d{\'e}riv{\'e}es partielles et leur signification
  physique},}\ }\href@noop {} {\bibfield  {journal} {\bibinfo  {journal}
  {Princeton university bulletin}\ }\textbf {\bibinfo {volume} {13}},\ \bibinfo
  {pages} {28} (\bibinfo {year} {1902})}\BibitemShut {NoStop}%
\bibitem [{\citenamefont {Epstein}(1969)}]{epstein1969}%
  \BibitemOpen
  \bibfield  {author} {\bibinfo {author} {\bibfnamefont {E.~S.}\ \bibnamefont
  {Epstein}},\ }\bibfield  {title} {\enquote {\bibinfo {title} {Stochastic
  dynamic prediction},}\ }\href@noop {} {\bibfield  {journal} {\bibinfo
  {journal} {Tellus}\ }\textbf {\bibinfo {volume} {21}},\ \bibinfo {pages}
  {739--759} (\bibinfo {year} {1969})}\BibitemShut {NoStop}%
\bibitem [{\citenamefont {Palmer}(2000)}]{palmer2000}%
  \BibitemOpen
  \bibfield  {author} {\bibinfo {author} {\bibfnamefont {T.~N.}\ \bibnamefont
  {Palmer}},\ }\bibfield  {title} {\enquote {\bibinfo {title} {Predicting
  uncertainty in forecasts of weather and climate},}\ }\href@noop {} {\bibfield
   {journal} {\bibinfo  {journal} {Reports on Progress in Physics}\ }\textbf
  {\bibinfo {volume} {63}},\ \bibinfo {pages} {71} (\bibinfo {year}
  {2000})}\BibitemShut {NoStop}%
\bibitem [{\citenamefont {Thompson}(1957)}]{thompson1957}%
  \BibitemOpen
  \bibfield  {author} {\bibinfo {author} {\bibfnamefont {Philip~Duncan}\
  \bibnamefont {Thompson}},\ }\bibfield  {title} {\enquote {\bibinfo {title}
  {Uncertainty of initial state as a factor in the predictability of large
  scale atmospheric flow patterns},}\ }\href@noop {} {\bibfield  {journal}
  {\bibinfo  {journal} {Tellus}\ }\textbf {\bibinfo {volume} {9}},\ \bibinfo
  {pages} {275--295} (\bibinfo {year} {1957})}\BibitemShut {NoStop}%
\bibitem [{\citenamefont {Bauer}\ \emph {et~al.}(2015)\citenamefont {Bauer},
  \citenamefont {Thorpe},\ and\ \citenamefont {Brunet}}]{bauer2015}%
  \BibitemOpen
  \bibfield  {author} {\bibinfo {author} {\bibfnamefont {P.}~\bibnamefont
  {Bauer}}, \bibinfo {author} {\bibfnamefont {A.}~\bibnamefont {Thorpe}}, \
  and\ \bibinfo {author} {\bibfnamefont {G.}~\bibnamefont {Brunet}},\
  }\bibfield  {title} {\enquote {\bibinfo {title} {The quiet revolution of
  numerical weather prediction},}\ }\href@noop {} {\bibfield  {journal}
  {\bibinfo  {journal} {Nature}\ }\textbf {\bibinfo {volume} {525}},\ \bibinfo
  {pages} {47--55} (\bibinfo {year} {2015})}\BibitemShut {NoStop}%
\bibitem [{\citenamefont {Taylor}\ and\ \citenamefont
  {Buizza}(2003)}]{taylor2003}%
  \BibitemOpen
  \bibfield  {author} {\bibinfo {author} {\bibfnamefont {James~W}\ \bibnamefont
  {Taylor}}\ and\ \bibinfo {author} {\bibfnamefont {Roberto}\ \bibnamefont
  {Buizza}},\ }\bibfield  {title} {\enquote {\bibinfo {title} {Using weather
  ensemble predictions in electricity demand forecasting},}\ }\href@noop {}
  {\bibfield  {journal} {\bibinfo  {journal} {International Journal of
  Forecasting}\ }\textbf {\bibinfo {volume} {19}},\ \bibinfo {pages} {57--70}
  (\bibinfo {year} {2003})}\BibitemShut {NoStop}%
\bibitem [{\citenamefont {Emory}\ \emph {et~al.}(2013)\citenamefont {Emory},
  \citenamefont {Larsson},\ and\ \citenamefont {Iaccarino}}]{emory1}%
  \BibitemOpen
  \bibfield  {author} {\bibinfo {author} {\bibfnamefont {M.}~\bibnamefont
  {Emory}}, \bibinfo {author} {\bibfnamefont {J.}~\bibnamefont {Larsson}}, \
  and\ \bibinfo {author} {\bibfnamefont {G.}~\bibnamefont {Iaccarino}},\
  }\bibfield  {title} {\enquote {\bibinfo {title} {Modeling of structural
  uncertainties in reynolds-averaged navier-stokes closures},}\ }\href@noop {}
  {\bibfield  {journal} {\bibinfo  {journal} {Physics of Fluids
  (1994-present)}\ }\textbf {\bibinfo {volume} {25}},\ \bibinfo {pages}
  {110822} (\bibinfo {year} {2013})}\BibitemShut {NoStop}%
\bibitem [{\citenamefont {Gorl{\'e}}\ and\ \citenamefont
  {Iaccarino}(2013)}]{gorle1}%
  \BibitemOpen
  \bibfield  {author} {\bibinfo {author} {\bibfnamefont {C.}~\bibnamefont
  {Gorl{\'e}}}\ and\ \bibinfo {author} {\bibfnamefont {G.}~\bibnamefont
  {Iaccarino}},\ }\bibfield  {title} {\enquote {\bibinfo {title} {A framework
  for epistemic uncertainty quantification of turbulent scalar flux models for
  reynolds-averaged navier-stokes simulations},}\ }\href@noop {} {\bibfield
  {journal} {\bibinfo  {journal} {Physics of Fluids (1994-present)}\ }\textbf
  {\bibinfo {volume} {25}},\ \bibinfo {pages} {055105} (\bibinfo {year}
  {2013})}\BibitemShut {NoStop}%
\bibitem [{\citenamefont {Gorl{\'e}}\ \emph {et~al.}(2015)\citenamefont
  {Gorl{\'e}}, \citenamefont {Garcia-Sanchez},\ and\ \citenamefont
  {Iaccarino}}]{gorle2}%
  \BibitemOpen
  \bibfield  {author} {\bibinfo {author} {\bibfnamefont {C.}~\bibnamefont
  {Gorl{\'e}}}, \bibinfo {author} {\bibfnamefont {C.}~\bibnamefont
  {Garcia-Sanchez}}, \ and\ \bibinfo {author} {\bibfnamefont {G.}~\bibnamefont
  {Iaccarino}},\ }\bibfield  {title} {\enquote {\bibinfo {title} {Quantifying
  inflow and rans turbulence model form uncertainties for wind engineering
  flows},}\ }\href@noop {} {\bibfield  {journal} {\bibinfo  {journal} {Journal
  of Wind Engineering and Industrial Aerodynamics}\ }\textbf {\bibinfo {volume}
  {144}},\ \bibinfo {pages} {202--212} (\bibinfo {year} {2015})}\BibitemShut
  {NoStop}%
\bibitem [{\citenamefont {Garc{\'\i}a-S{\'a}nchez}\ \emph
  {et~al.}(2014)\citenamefont {Garc{\'\i}a-S{\'a}nchez}, \citenamefont
  {Philips},\ and\ \citenamefont {Gorl{\'e}}}]{gorle3}%
  \BibitemOpen
  \bibfield  {author} {\bibinfo {author} {\bibfnamefont {C.}~\bibnamefont
  {Garc{\'\i}a-S{\'a}nchez}}, \bibinfo {author} {\bibfnamefont {D.~A.}\
  \bibnamefont {Philips}}, \ and\ \bibinfo {author} {\bibfnamefont
  {C.}~\bibnamefont {Gorl{\'e}}},\ }\bibfield  {title} {\enquote {\bibinfo
  {title} {Quantifying inflow uncertainties for cfd simulations of the flow in
  downtown oklahoma city},}\ }\href@noop {} {\bibfield  {journal} {\bibinfo
  {journal} {Building and Environment}\ }\textbf {\bibinfo {volume} {78}},\
  \bibinfo {pages} {118--129} (\bibinfo {year} {2014})}\BibitemShut {NoStop}%
\bibitem [{\citenamefont {Xiao}\ \emph {et~al.}(2016)\citenamefont {Xiao},
  \citenamefont {Wu}, \citenamefont {Wang},\ and\ \citenamefont {Roy}}]{xiao}%
  \BibitemOpen
  \bibfield  {author} {\bibinfo {author} {\bibfnamefont {H.}~\bibnamefont
  {Xiao}}, \bibinfo {author} {\bibfnamefont {J.-L.}\ \bibnamefont {Wu}},
  \bibinfo {author} {\bibfnamefont {R.}~\bibnamefont {Wang}}, \ and\ \bibinfo
  {author} {\bibfnamefont {C.J.}\ \bibnamefont {Roy}},\ }\bibfield  {title}
  {\enquote {\bibinfo {title} {Quantifying and reducing model form
  uncertainties in reynolds-averaged navier-stokes simulations: A data-driven,
  physics-informed approach},}\ }\href@noop {} {\bibfield  {journal} {\bibinfo
  {journal} {Journal of Computational Physics}\ }\textbf {\bibinfo {volume}
  {324}},\ \bibinfo {pages} {115--136} (\bibinfo {year} {2016})}\BibitemShut
  {NoStop}%
\bibitem [{\citenamefont {Mishra}\ \emph
  {et~al.}(2016{\natexlab{a}})\citenamefont {Mishra}, \citenamefont
  {Iaccarino},\ and\ \citenamefont {Duraisamy}}]{mishra2016}%
  \BibitemOpen
  \bibfield  {author} {\bibinfo {author} {\bibfnamefont {Aashwin~A}\
  \bibnamefont {Mishra}}, \bibinfo {author} {\bibfnamefont {Gianluca}\
  \bibnamefont {Iaccarino}}, \ and\ \bibinfo {author} {\bibfnamefont {Karthik}\
  \bibnamefont {Duraisamy}},\ }\bibfield  {title} {\enquote {\bibinfo {title}
  {Sensitivity of flow evolution on turbulence structure},}\ }\href@noop {}
  {\bibfield  {journal} {\bibinfo  {journal} {Physical Review Fluids}\ }\textbf
  {\bibinfo {volume} {1}},\ \bibinfo {pages} {052402} (\bibinfo {year}
  {2016}{\natexlab{a}})}\BibitemShut {NoStop}%
\bibitem [{\citenamefont {Mishra}\ \emph
  {et~al.}(2016{\natexlab{b}})\citenamefont {Mishra}, \citenamefont
  {Iaccarino},\ and\ \citenamefont {Duraisamy}}]{mishra5}%
  \BibitemOpen
  \bibfield  {author} {\bibinfo {author} {\bibfnamefont {A.~A.}\ \bibnamefont
  {Mishra}}, \bibinfo {author} {\bibfnamefont {G.}~\bibnamefont {Iaccarino}}, \
  and\ \bibinfo {author} {\bibfnamefont {K.}~\bibnamefont {Duraisamy}},\
  }\bibfield  {title} {\enquote {\bibinfo {title} {Epistemic uncertainty in
  statistical markovian turbulence models},}\ }\href@noop {} {\bibfield
  {journal} {\bibinfo  {journal} {Center for Turbulence Research Annual
  Research Briefs}\ ,\ \bibinfo {pages} {183--195}} (\bibinfo {year}
  {2016}{\natexlab{b}})}\BibitemShut {NoStop}%
\bibitem [{\citenamefont {Williams}(2005)}]{williams2005}%
  \BibitemOpen
  \bibfield  {author} {\bibinfo {author} {\bibfnamefont {Paul~D}\ \bibnamefont
  {Williams}},\ }\bibfield  {title} {\enquote {\bibinfo {title} {Modelling
  climate change: the role of unresolved processes},}\ }\href@noop {}
  {\bibfield  {journal} {\bibinfo  {journal} {Philosophical Transactions of the
  Royal Society of London A: Mathematical, Physical and Engineering Sciences}\
  }\textbf {\bibinfo {volume} {363}},\ \bibinfo {pages} {2931--2946} (\bibinfo
  {year} {2005})}\BibitemShut {NoStop}%
\bibitem [{\citenamefont {Toth}\ and\ \citenamefont {Kalnay}(1997)}]{zoltan}%
  \BibitemOpen
  \bibfield  {author} {\bibinfo {author} {\bibfnamefont {Z.}~\bibnamefont
  {Toth}}\ and\ \bibinfo {author} {\bibfnamefont {E.}~\bibnamefont {Kalnay}},\
  }\bibfield  {title} {\enquote {\bibinfo {title} {Ensemble forecasting at ncep
  and the breeding method},}\ }\href@noop {} {\bibfield  {journal} {\bibinfo
  {journal} {Monthly Weather Review}\ }\textbf {\bibinfo {volume} {125}},\
  \bibinfo {pages} {3297--3319} (\bibinfo {year} {1997})}\BibitemShut {NoStop}%
\bibitem [{\citenamefont {Kassinos}\ and\ \citenamefont
  {Reynolds}(1996)}]{iprm2}%
  \BibitemOpen
  \bibfield  {author} {\bibinfo {author} {\bibfnamefont {S.~C.}\ \bibnamefont
  {Kassinos}}\ and\ \bibinfo {author} {\bibfnamefont {W.~C.}\ \bibnamefont
  {Reynolds}},\ }\bibfield  {title} {\enquote {\bibinfo {title} {A particle
  representation model for the deformation of homogeneous turbulence},}\
  }\href@noop {} {\bibfield  {journal} {\bibinfo  {journal} {Annual Research
  Briefs-1996}\ ,\ \bibinfo {pages} {31}} (\bibinfo {year} {1996})}\BibitemShut
  {NoStop}%
\bibitem [{\citenamefont {Kassinos}\ and\ \citenamefont
  {Reynolds}(1997)}]{iprm}%
  \BibitemOpen
  \bibfield  {author} {\bibinfo {author} {\bibfnamefont {S.~C.}\ \bibnamefont
  {Kassinos}}\ and\ \bibinfo {author} {\bibfnamefont {W.~C.}\ \bibnamefont
  {Reynolds}},\ }\bibfield  {title} {\enquote {\bibinfo {title} {Advances in
  structure-based turbulence modeling},}\ }\href@noop {} {\bibfield  {journal}
  {\bibinfo  {journal} {Annual Research Briefs 1997}\ ,\ \bibinfo {pages}
  {179--193}} (\bibinfo {year} {1997})}\BibitemShut {NoStop}%
\bibitem [{\citenamefont {Kassinos}\ \emph {et~al.}(2001)\citenamefont
  {Kassinos}, \citenamefont {Reynolds},\ and\ \citenamefont {Rogers}}]{krr}%
  \BibitemOpen
  \bibfield  {author} {\bibinfo {author} {\bibfnamefont {S.~C.}\ \bibnamefont
  {Kassinos}}, \bibinfo {author} {\bibfnamefont {W.~C.}\ \bibnamefont
  {Reynolds}}, \ and\ \bibinfo {author} {\bibfnamefont {M.~M.}\ \bibnamefont
  {Rogers}},\ }\bibfield  {title} {\enquote {\bibinfo {title} {One-point
  turbulence structure tensors},}\ }\href@noop {} {\bibfield  {journal}
  {\bibinfo  {journal} {Journal of Fluid Mechanics}\ }\textbf {\bibinfo
  {volume} {428}},\ \bibinfo {pages} {213--248} (\bibinfo {year}
  {2001})}\BibitemShut {NoStop}%
\bibitem [{\citenamefont {Campos}(2016)}]{campos}%
  \BibitemOpen
  \bibfield  {author} {\bibinfo {author} {\bibfnamefont {A.}~\bibnamefont
  {Campos}},\ }\emph {\bibinfo {title} {Advances in structure-based modeling of
  turbulent flows}},\ \href@noop {} {Ph.D. thesis},\ \bibinfo  {school}
  {Stanford University} (\bibinfo {year} {2016})\BibitemShut {NoStop}%
\bibitem [{\citenamefont {Toth}\ and\ \citenamefont {Kalnay}(1993)}]{toth}%
  \BibitemOpen
  \bibfield  {author} {\bibinfo {author} {\bibfnamefont {Zoltan}\ \bibnamefont
  {Toth}}\ and\ \bibinfo {author} {\bibfnamefont {Eugenia}\ \bibnamefont
  {Kalnay}},\ }\bibfield  {title} {\enquote {\bibinfo {title} {Ensemble
  forecasting at nmc: The generation of perturbations},}\ }\href@noop {}
  {\bibfield  {journal} {\bibinfo  {journal} {Bulletin of the american
  meteorological society}\ }\textbf {\bibinfo {volume} {74}},\ \bibinfo {pages}
  {2317--2330} (\bibinfo {year} {1993})}\BibitemShut {NoStop}%
\bibitem [{\citenamefont {Leith}(1974)}]{leith1974}%
  \BibitemOpen
  \bibfield  {author} {\bibinfo {author} {\bibfnamefont {C.~E.}\ \bibnamefont
  {Leith}},\ }\bibfield  {title} {\enquote {\bibinfo {title} {Theoretical skill
  of monte carlo forecasts},}\ }\href@noop {} {\bibfield  {journal} {\bibinfo
  {journal} {Monthly Weather Review}\ }\textbf {\bibinfo {volume} {102}},\
  \bibinfo {pages} {409--418} (\bibinfo {year} {1974})}\BibitemShut {NoStop}%
\bibitem [{\citenamefont {Guiasu}\ and\ \citenamefont
  {Shenitzer}(1985)}]{maxent}%
  \BibitemOpen
  \bibfield  {author} {\bibinfo {author} {\bibfnamefont {Silviu}\ \bibnamefont
  {Guiasu}}\ and\ \bibinfo {author} {\bibfnamefont {Abe}\ \bibnamefont
  {Shenitzer}},\ }\bibfield  {title} {\enquote {\bibinfo {title} {The principle
  of maximum entropy},}\ }\href@noop {} {\bibfield  {journal} {\bibinfo
  {journal} {The mathematical intelligencer}\ }\textbf {\bibinfo {volume}
  {7}},\ \bibinfo {pages} {42--48} (\bibinfo {year} {1985})}\BibitemShut
  {NoStop}%
\bibitem [{\citenamefont {Soize}(2008)}]{soize2008}%
  \BibitemOpen
  \bibfield  {author} {\bibinfo {author} {\bibfnamefont {Christian}\
  \bibnamefont {Soize}},\ }\bibfield  {title} {\enquote {\bibinfo {title}
  {Construction of probability distributions in high dimension using the
  maximum entropy principle: Applications to stochastic processes, random
  fields and random matrices},}\ }\href@noop {} {\bibfield  {journal} {\bibinfo
   {journal} {International Journal for Numerical Methods in Engineering}\
  }\textbf {\bibinfo {volume} {76}},\ \bibinfo {pages} {1583--1611} (\bibinfo
  {year} {2008})}\BibitemShut {NoStop}%
\bibitem [{\citenamefont {Baggett}(1997)}]{baggett}%
  \BibitemOpen
  \bibfield  {author} {\bibinfo {author} {\bibfnamefont {J.~S.}\ \bibnamefont
  {Baggett}},\ }\bibfield  {title} {\enquote {\bibinfo {title} {Some modeling
  requirements for wall models in large eddy simulation},}\ }\href@noop {}
  {\bibfield  {journal} {\bibinfo  {journal} {Annual Research Briefs}\ ,\
  \bibinfo {pages} {123--134}} (\bibinfo {year} {1997})}\BibitemShut {NoStop}%
\bibitem [{\citenamefont {Nicoud}\ \emph {et~al.}(1998)\citenamefont {Nicoud},
  \citenamefont {Winckelmans}, \citenamefont {Carati}, \citenamefont
  {Baggett},\ and\ \citenamefont {Cabot}}]{cabot}%
  \BibitemOpen
  \bibfield  {author} {\bibinfo {author} {\bibfnamefont {F.}~\bibnamefont
  {Nicoud}}, \bibinfo {author} {\bibfnamefont {G.}~\bibnamefont {Winckelmans}},
  \bibinfo {author} {\bibfnamefont {D.}~\bibnamefont {Carati}}, \bibinfo
  {author} {\bibfnamefont {J.}~\bibnamefont {Baggett}}, \ and\ \bibinfo
  {author} {\bibfnamefont {W.}~\bibnamefont {Cabot}},\ }\bibfield  {title}
  {\enquote {\bibinfo {title} {Boundary conditions for les away from the
  wall},}\ }\href@noop {} {\bibfield  {journal} {\bibinfo  {journal}
  {Proceedings of the CTR Summer Program}\ ,\ \bibinfo {pages} {413--422}}
  (\bibinfo {year} {1998})}\BibitemShut {NoStop}%
\bibitem [{\citenamefont {Isaza}\ and\ \citenamefont
  {Collins}(2009)}]{isazancollins}%
  \BibitemOpen
  \bibfield  {author} {\bibinfo {author} {\bibfnamefont {Juan~C}\ \bibnamefont
  {Isaza}}\ and\ \bibinfo {author} {\bibfnamefont {Lance~R}\ \bibnamefont
  {Collins}},\ }\bibfield  {title} {\enquote {\bibinfo {title} {On the
  asymptotic behaviour of large-scale turbulence in homogeneous shear flow},}\
  }\href@noop {} {\bibfield  {journal} {\bibinfo  {journal} {Journal of Fluid
  Mechanics}\ }\textbf {\bibinfo {volume} {637}},\ \bibinfo {pages} {213--239}
  (\bibinfo {year} {2009})}\BibitemShut {NoStop}%
\bibitem [{\citenamefont {Mishra}\ and\ \citenamefont
  {Girimaji}(2013)}]{mishra2}%
  \BibitemOpen
  \bibfield  {author} {\bibinfo {author} {\bibfnamefont {A.~A.}\ \bibnamefont
  {Mishra}}\ and\ \bibinfo {author} {\bibfnamefont {S.~S.}\ \bibnamefont
  {Girimaji}},\ }\bibfield  {title} {\enquote {\bibinfo {title} {Intercomponent
  energy transfer in incompressible homogeneous turbulence: multi-point physics
  and amenability to one-point closures},}\ }\href@noop {} {\bibfield
  {journal} {\bibinfo  {journal} {Journal of Fluid Mechanics}\ }\textbf
  {\bibinfo {volume} {731}},\ \bibinfo {pages} {639--681} (\bibinfo {year}
  {2013})}\BibitemShut {NoStop}%
\bibitem [{\citenamefont {Mishra}\ and\ \citenamefont
  {Girimaji}(2014)}]{mishra3}%
  \BibitemOpen
  \bibfield  {author} {\bibinfo {author} {\bibfnamefont {A.~A.}\ \bibnamefont
  {Mishra}}\ and\ \bibinfo {author} {\bibfnamefont {S.~S.}\ \bibnamefont
  {Girimaji}},\ }\bibfield  {title} {\enquote {\bibinfo {title} {On the
  realizability of pressure--strain closures},}\ }\href@noop {} {\bibfield
  {journal} {\bibinfo  {journal} {Journal of Fluid Mechanics}\ }\textbf
  {\bibinfo {volume} {755}},\ \bibinfo {pages} {535--560} (\bibinfo {year}
  {2014})}\BibitemShut {NoStop}%
\bibitem [{\citenamefont {Magnus}\ and\ \citenamefont {Winkler}(2013)}]{hills}%
  \BibitemOpen
  \bibfield  {author} {\bibinfo {author} {\bibfnamefont {W.}~\bibnamefont
  {Magnus}}\ and\ \bibinfo {author} {\bibfnamefont {S.}~\bibnamefont
  {Winkler}},\ }\href@noop {} {\emph {\bibinfo {title} {Hill's equation}}}\
  (\bibinfo  {publisher} {Courier Corporation},\ \bibinfo {year}
  {2013})\BibitemShut {NoStop}%
\end{thebibliography}%
\end{document}